    \documentclass{aa}  

    \usepackage{graphicx}
    \usepackage{txfonts}
    \usepackage[colorlinks = true, allcolors=blue]{hyperref}
    \makeatletter
    \renewcommand*\aa@pageof{, page \thepage{} of \pageref*{LastPage}}
    \makeatother
    \usepackage{booktabs}
    \usepackage{soul}
    \usepackage{float}
\newcommand{\orcidlink}[1]{\protect\href{https://orcid.org/#1}{\protect\includegraphics[width=8pt]{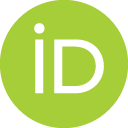}}}    
    
\begin{document}

\title{The effect of the environment-dependent stellar initial mass function on the photometric properties of star-forming galaxies}
\titlerunning{Star-forming galaxies in the IGIMF theory}
\authorrunning{Haslbauer, Yan, Jerabkova, Gjergo, Zonoozi, and Kroupa}
       
\author{Moritz Haslbauer\orcidlink{0000-0002-5101-6366}\inst{1},
              Zhiqiang Yan\orcidlink{0000-0001-7395-1198}\inst{2,3},
              Tereza Jerabkova\orcidlink{0000-0002-1251-9905}\inst{4},
              Eda Gjergo\orcidlink{0000-0002-7440-1080}\inst{2,3},
              Pavel Kroupa\orcidlink{0000-0002-7301-3377}\inst{1,5}
              \and 
              Akram Hasani Zonoozi\orcidlink{0000-0002-0322-9957}\inst{6} 
              }
    
\institute{
            Helmholtz-Institut für Strahlen- und Kernphysik, Universität Bonn, Nussallee 14-16, D-53115 Bonn, Germany\\
            \email{mhaslbauer@astro.uni-bonn.de}
            \and
            School of Astronomy and Space Science, Nanjing University, Nanjing 210093, People’s Republic of China \\
            \email{yan@nju.edu.cn}
            \and
            Key Laboratory of Modern Astronomy and Astrophysics (Nanjing University), Ministry of Education, Nanjing 210093, People’s Republic of China
            \and
            European Southern Observatory, Karl-Schwarzschild-Straße 2, 85748 Garching bei M\"unchen, Germany\\
            \email{tereza.jerabkova@eso.org}
            \and
            Astronomical Institute, Faculty of Mathematics and Physics, Charles University, V Hole\v{s}ovi\v{c}k\'ach 2, CZ-180 00 Praha 8, Czech Republic
            \and
            Department of Physics, Institute for Advanced Studies in Basic Sciences (IASBS), PO Box 11365-9161, Zanjan, Iran 
            \\
}
    
\date{Accepted April 30, 2024}

     
\abstract
    {Observational estimates of galaxy properties, including mass and star formation rates (SFRs), rely on the inherent galaxy-wide initial mass function (gwIMF), which systematically varies with the global SFR and metallicity, as proposed by the integrated-galactic IMF (IGIMF) theory and supported by empirical evidence.}
    {We aim to evaluate the influence of the variable gwIMF on various galaxy properties, encompassing the Ks-, K$_{\mathrm{3.6}}$-, and V-band stellar mass-to-light ratio, SFR--luminosity relation, gas depletion timescale, and stellar mass buildup timescale of local star-forming galaxies.}
    {We incorporate PARSEC and COLIBRI stellar isochrones into the GalIMF code, a galaxy chemical evolution (GCE) model featuring real-time updates of environment-dependent gwIMFs. This newly developed photometric GalIMF (photGalIMF) code allows the calculation of photometric properties for galaxies with diverse stellar populations. Subsequently, we analyze observed luminosities and metallicities of local star-forming galaxies to deduce their stellar masses assuming that they have constant SFRs over 13.6 Gyr. We also compute SFR--H$\alpha$ luminosity relations for varying stellar metallicities using a separate stellar population synthesis code based on P\'{E}GASE.}
    {Comparing the IGIMF theory to the canonical universal IMF, our analysis reveals that estimates of the stellar masses and SFRs for local star-forming galaxies differ by factors of approximately 2 and 10, respectively. This disparity yields a well-defined galaxy main sequence extending to dwarf galaxies. The computed gas-depletion timescale increases with gas mass, implying lower star formation efficiencies in more massive galaxies, possibly due to stronger feedback regulation, aligning with theoretical expectations. Additionally, the characteristic stellar mass buildup timescale increases with stellar mass, indicating that massive disk galaxies initiate star formation earlier than their low-mass counterparts.}
    {The photGalIMF code enables self-consistent computations of galactic photometry, self-consistently with GCE modelling within the context of an environment-dependent gwIMF. Utilizing Ks-band and H$\alpha$ luminosities of galaxies, the outcomes include galaxy mass, SFR, and fitting functions for the SFR correction factor.}

    
\maketitle
    %

\section{Introduction}
\label{sec:Introduction}

Galaxies in the observed distant and local Universe constrain various galaxy evolution and cosmological models, which require precise measurements and a correct interpretation of the measured observables. The inferred physical properties of galaxies are not only sensitive to the underlying cosmological framework \citep{BalakrishnaSubramani_2019, Haslbauer_2022b} but also to the behaviour of baryonic physics (e.g. \citealt{2015MNRAS.446..521S}, \citealt{Ludlow_2017}, \citealt{2018MNRAS.480..800H}, \citealt{Li_2020}, and \citealt{2022MNRAS.512..199K}. Interestingly, the dependency on baryonic physics is much less pronounced when Milgromian gravitation is applied, \citealt{Wittenburg_2020, Eappen_2022, Nagesh_2023}). 

A fundamental parameter in galaxy evolution is the initial stellar mass function (IMF) which describes the mass distribution of newly formed stars \citep{Hopkins_2018}. The IMF together with the star formation history (SFH) determine the stellar population of systems, the masses in living stars and remnants, and affect the matter cycle and chemical enrichment. Deducing the SFRs of unresolved stellar systems requires measurements of SFR indicators which are available, for example, for the X-ray, ultra-violet, optical, infrared, and the radio regime \citep[e.g.][and references therein]{Kennicutt_1983, Donas_1984, Buat_1989, David_1992, Condon_1992, Kennicutt_1998, Yun_2001, Kewley_2002, Brinchmann_2004, Moustakas_2006, Rieke_2009, Kennicutt_2012, Calzetti_2013, Brown_2017, Mahajan_2019}. Crucially, converting these tracers to a SFR depends on the shape of the IMF \citep{Kroupa_Jerabkova_2021}. A strong variation of the IMF can alter galaxy mass estimates by more than a factor of ten \citep{2023arXiv231006781W}. 
Therefore, the properties and functional dependencies of the IMF are of great importance in interpreting the observations of stellar systems. 

It has been commonly assumed in the past that the IMF is invariant (known as the canonical IMF, see \citealt{Kroupa_2002} that highlights the uniformity of IMFs of nearby star clusters) meaning that the mass distribution of stars is independent of the star-forming environment. However, increasingly observations suggest that the IMF varies inside our Galaxy (\citealt{Matteucci_1994, Vazdekis_1997, Kroupa_2002, Marks_2012,2014MNRAS.444.1957D,Dib_2017,2023ApJ...959...88D,Li_2023}) and outside if as assessed in systems ranging from dense star-formation regions \citep{Dabringhausen_2009,Dabringhausen_2012,Schneider_2018,Zhang_2018} to dwarf galaxies \citep{Meurer_2009,Lee_2009,Watts_2018} and from very metal-poor satellite galaxies \citep{Geha_2013,Gennaro_2018,Yan_2020,Mucciarelli_2021} to galaxies with about Solar metallicity \citep{Gunawardhana_2011,Parikh_2018,Zhou_2019,MartinNavarro_2019,Smith_2020,vanDokkum_2021}. The JWST observation of high-redshift galaxy spectra is also consistent with a top-heavy IMF \citep{2023arXiv231102051C}.

The integrated galactic IMF (IGIMF) theory, introduced by \citet{Kroupa_2003}, can largely account for the above variety of observations.
It is a mathematical framework to calculate the galaxy-wide IMF (gwIMF) by adding up the stellar IMFs of all embedded clusters within a galaxy. Note that from here on we differentiate between the IMF which is the result of one star-formation event as an embedded star cluster in a molecular cloud and the gwIMF. Both have the same shape only if the former is a scale-invariant probability density distribution function \citep{Kroupa_Jerabkova_2021}. Qualitatively, the IGIMF theory relies on the empirical relations of star formation and has been constantly further constrained \citep{Kroupa_2013,2013MNRAS.435.2274W,2013MNRAS.434...84W,2013MNRAS.436.3309W,2014MNRAS.441.3348W,2015A&A...582A..93S,Yan_2017,Yan_2019a,Yan_2020,Yan_2023,Jerabkova_2018,Fontanot_2018,2023arXiv231112932F,2018A&A...614A..43D,Zonoozi_2019}. 

According to the IGIMF theory, the gwIMF systematically varies with the properties of the star-forming environment such as the global SFR, $\psi$, and averaged gas-phase metallicity of a galaxy. For example, the gwIMF for galaxies with a higher SFR than the Milky Way ($\psi \ga \mathrm{few} \,\rm{M_{\odot}\,yr^{-1}}$) becomes top-heavy \citep{2011MNRAS.412..979W,Gunawardhana_2011,Zhang_2018}, meaning that the formation of more massive stars is favoured compared to the canonical IMF, and the gwIMF for low-SFR galaxies ($\psi < 1\,\rm{M_{\odot}\,yr^{-1}}$) becomes top-light \citep{Lee_2009,Yan_2020,Mucciarelli_2021}, resulting in a steeper gwIMF slope for stars above $1~M_\odot$ relative to the canonical IMF of the Milky Way. The metallicity also correlates with the slope of the IMF for both massive \citep{Marks_2012} and low-mass stars \citep[and references therein]{Yan24}. For metal-poor and metal-rich stellar populations, their IMFs become bottom-light (relatively fewer low-mass stars) and bottom-heavy (steeper IMF slope for low-mass stars), respectively \citep[see e.g. Fig.~2 of][and Section~\ref{subsec:IGIMF theory}]{Jerabkova_2018}. The IMF is likely to be affected by more physical parameters, such as cosmic ray density \citep{Papadopoulos_2010,Fontanot_2018} and temperature \citep{Sneppen_2022}, but these initial conditions are difficult to measure and can be affected by the SFR density. 
With only two environmental inputs, the galactic SFR and metallicity, the IGIMF model has successfully explained several observations over the last two decades. For example, the mass--metallicity relation \citep[][]{Koeppen_2007,Yan_2019b,Yan_2021}, the [$\alpha$/Fe] relation in early-type galaxies \citep[UFDs;][]{Recchi_2009,Yan_2019b,Yan_2021}, the chemical evolution of ultra-faint dwarf galaxies \citep{Yan_2020,Mucciarelli_2021}, the IMF of UFD galaxies \citep{2022A&A...666A.113D}, the UV/H$\alpha$ flux ratio of galaxies \citep[][]{Lee_2009,PflammAltenburg_2009a,Yan_2017}, the radial H$\alpha$ cut-off in disk galaxies \citep[][]{PflammAltenburg_2008}, and the mass correlation between spheroidal galaxies and their hosted supermassive black holes \citep[][]{Kroupa_2020_SMBH}. 

We note that \citet{2022A&A...666A.113D} argues for a possible random variation of the IMF from one star-forming region to another. This is an interesting possibility, but for the time being here we consider that the IMF is optimally sampled (i.e. there are no random variations) from an underlying parent function which strictly only depends on the metallicity and density of the embedded-cluster forming gas cloud core, following \citet{Yan_2017}. This gives us complete analytical predictability and thus allows detailed controlled experiments ranging from the embedded cluster scale to the scale of whole galaxies. The possible random variations proposed by \citet{2022A&A...666A.113D} will be studied in the future.

The implications of the systematic variation of the gwIMF given by the IGIMF theory are significant. For example, the number of ionizing photons depends on the shape of the gwIMF such that the relation between the SFR and $\mathrm{H\alpha}$ luminosity, $L_{\mathrm{H\alpha}}$, of a galaxy becomes non-linear for a varying IMF \citep{PflammAltenburg_2009a,Jerabkova_2018} violating therewith the Kennicutt law \citep{Kennicutt_1998}. Galaxies with $L_{\mathrm{H\alpha}} > 10^{41}\,\rm{ergs/s}$ ($L_{\mathrm{H\alpha}} < 10^{41}\,\rm{ergs/s}$) have SFRs lower (higher) than those given by the Kennicutt law. This affects the properties of star-forming galaxies such as the main sequence of star-forming galaxies, which is a tight correlation between the SFRs and stellar masses of galaxies \citep[see e.g. Fig.~8 of][]{Speagle_2014}. Applying an IGIMF-corrected SFR--H$\alpha$ luminosity relation \citep[see Fig.~7 of][]{Jerabkova_2018}, the slope of the main sequence becomes flatter \citep{Kroupa_Jerabkova_2021}. Another consequence of the variable IMF is for the estimation of the gas depletion timescale, $\tau_{\mathrm{gas}} = M_{\mathrm{gas}}/\psi$, and stellar-mass build-up timescale, $\tau_\star = M_\star/\psi$. \citet{PflammAltenburg_2009b} calculated these timescales for 200 nearby galaxies in the canonical IMF and IGIMF context. They demonstrated that in the IGIMF framework, the star formation efficiency (SFE), $\tau_{\mathrm{gas}}^{-1}$, remains almost constant, while massive disk galaxies have much higher $\tau_{\mathrm{gas}}^{-1}$ than dwarf galaxies for a canonical IMF. Moreover, \citet{PflammAltenburg_2009b} find that $\tau_\star$ increases with stellar mass in the IGIMF framework, implying that SFR may have been increasing slightly with time for dwarf galaxies but not for large disk galaxies. This is in agreement with observations \citep{Fontanot_2009} and in contrast to the canonical IMF expectations. 

The above studies mostly take into account the effect of the IMF variation on the estimation of the galactic SFR but not on the stellar mass. This simplification is reasonable because the SFR correction is much more significant than the stellar mass correction but still is not fully self-consistent if the latter is not also corrected.
The effect of the IMF variation on the determination of the galaxy mass in stars and their remnants is more complicated than the SFR estimation because it involves the SFH and chemical enrichment history. A detailed galaxy evolution simulation incorporating an environment-dependent IMF, as developed in \citet{Yan_2019a}, is necessary (cf. \citealt{2023arXiv231212109R}).

This work expands the GalIMF galaxy chemical evolution code \citep{Yan_2019a} to include galaxy photometry, that is, the stellar isochrones provided by the PAdova and TRieste Stellar Evolution Code (PARSEC, \citealt{Bressan_2012}) and COLIBRI code \citep{Marigo_2013}. This expanded public code is available under the name ``photometric GalIMF" (photGalIMF, Section~\ref{subsubsec:stellar evolution}). This code enables, for the first time, to study photometric properties of galaxies in self-consistent chemical evolution models with an environment-dependent gwIMF. Additionally, we integrate the methods developed in \citet{Jerabkova_2018} to obtain the SFR--H$\alpha$-luminosity relation, as well as \citet{PflammAltenburg_2009b} to treat galactic gas depletion timescales. We then correct both the SFRs and the stellar masses of galaxies located in the Local Cosmological Volume \citep[LV,][]{Karachentsev_2004,Karachentsev_2013}.
\begin{figure*}[!hbt]
\includegraphics[width=\hsize]{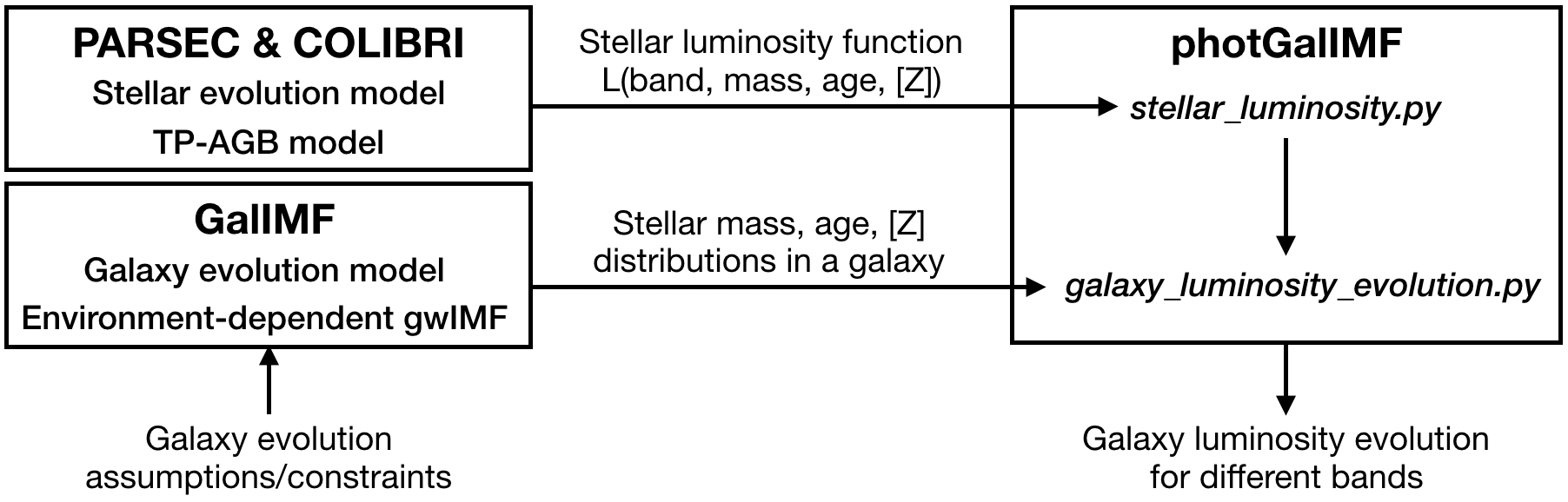}
\caption{Flowchart of the photGalIMF code, original to the present work, which couples the galaxy chemical evolution code GalIMF \citep{Yan_2019a} with stellar evolution models to calculate photometric properties of galaxies. The GalIMF code calculates the stellar population of a galaxy for given galaxy evolution assumptions (Section~\ref{subsubsec:Chemical enrichtment}) and a variable IMF that depends on the environment at each timestep. Throughout this work, we adopt either an invariant canonical IMF as the gwIMF (the ``canonical gwIMF'') or the ``IGIMF-2021'' gwIMF that depends on galactic SFR and gas metallicity. The metallicity, mass, and age distributions of stars are then combined with the photometric module to calculate the luminosity for selected photometric bands. See details in Section~\ref{subsubsec:stellar evolution}.}
\label{figure_photGalIMF_flowchart}
\end{figure*}

The paper is organized as follows: Section~\ref{sec: Methods} introduces our model and assumptions to calculate the stellar mass-to-light ratio ($M_\star/L$\, which includes the stellar remnants) in different photometric bands and the SFR--H$\alpha$-luminosity relation. Section~\ref{sec:observations} describes the observational constraints of our model and a sample of 603 star-forming galaxies in the LV ($< 11 {\rm Mpc}$). Section~\ref{sec: Results} presents the evolution of $M_\star/L$, the stellar-mass--SFR relation, the gas depletion timescale, and stellar-mass buildup times of local star-forming galaxies in the invariant canonical IMF and the IGIMF frameworks. The results are discussed in Section~\ref{sec: Discussion} which is followed by an outlook and a conclusion in Section~\ref{sec: Conclusion}. Throughout the analysis, we assume an age of the Universe of $\tau_{\mathrm{h}} = 13.8$~Gyr \citep[table~4 of][]{Planck_2016_IllustrisTNG}. 

\section{Methods}
\label{sec: Methods}
As introduced in Section~\ref{sec:Introduction}, the environment-dependent empirical IMF variation described by the IGIMF theory is both well-developed and rigorously tested. Here, we state the physical assumptions and mathematical framework of the IGIMF theory in Section~\ref{subsec:IGIMF theory}, providing qualitative discussions on the functional dependence of the gwIMFs on the global SFR and the metallicity of different IGIMF formulations. A more comprehensive description and the observational constraints of the IGIMF theory can be found, for example, in \citet{Jerabkova_2018} and \citet{Yan_2021,Yan24}. 
Next, we summarize our chemical evolution model, which is necessary to obtain the evolution of stellar ages, masses, and metallicity in Section~\ref{subsubsec:Chemical enrichtment}.
The assumption on the SFH with a constant SFR is motivated in Section~\ref{subsubsec:Star formation histories}.
We then present in Section~\ref{subsubsec:stellar evolution} the new photGalIMF code, which tracks the stellar luminosity evolution of synthetic photometric bands. It is an extension to the chemical evolution code GalIMF \citep{Yan_2019a}, which itself is built upon the IGIMF module \citep{Yan_2017}. 
A schematic representation of the code structure is shown in Fig.~\ref{figure_photGalIMF_flowchart}.
Finally, the SFR--H$\alpha$ luminosity relations for different metallicities are provided in Section~\ref{subsec: SFR-Halpha relation}.

\subsection{The IGIMF theory}\label{subsec:IGIMF theory}

\begin{table*}[!hbt]
 \caption{\label{t7} \textbf{Overview of different IGIMF formulations.}}
 \centering
	\begin{tabular}{l|ll|ll|ll}
	\hline
   Model & $\alpha_{1}$ \& $\alpha_{2}$ & Reference & $\alpha_{3}$ & Reference & $\beta$ & Reference \\ \hline  
   
   IGIMF-2003 & Fixed & \citet{Kroupa_2001} & Fixed & \citet{Kroupa_2001} & Fixed & \citet{2002AJ....124.1393L} \\
   
   IGIMF-2011 & Fixed & \citet{Kroupa_2001} & Varies & \citet{2011MNRAS.412..979W} & Varies & \citet{2011MNRAS.412..979W} \\
   
   IGIMF-2017 & Fixed & \citet{Kroupa_2001} & Varies & \citet{Marks_2012} & Varies & \citet{2013MNRAS.436.3309W} \\
   
   IGIMF-2018 & $0.5 {\rm \cdot [Fe/H]}$ & \citet{Kroupa_2002} & Varies & \citet{Marks_2012} & Varies & \citet{2013MNRAS.436.3309W} \\
   
   IGIMF-2020 & $35\cdot({\rm Z}-0.02)$ & \citet{Yan_2020} & Varies & \citet{Marks_2012} & Varies & \citet{2013MNRAS.436.3309W} \\
   
   IGIMF-2021 & $63\cdot({\rm Z}-0.0142)$ & \citet{Yan_2021} & Varies & \citet{Marks_2012} & Varies & \citet{2013MNRAS.436.3309W} \\
   
   IGIMF-2024 & $80\cdot({\rm Z}-0.0113)$ & \citet{Yan24} & Varies & \citet{Marks_2012} & Varies & \citet{2013MNRAS.436.3309W} \\ \hline
	\end{tabular}
	\tablefoot{Variation of the power-law indices of the stellar IMF (equations~\ref{eq: IMF}, \ref{eq:alpha_12}, and \ref{eq:alpha_3}) applied in different models, including IGIMF-2003 from \citet{Kroupa_2003}, IGIMF-2011 from \citet{2011MNRAS.412..979W}, IGIMF-2017 from \citet{Yan_2017}, IGIMF-2018 from \citet{Jerabkova_2018}, IGIMF-2020 from \citet{Yan_2020}, and IGIMF-2021 from \citet{Yan_2021}.}
 \label{tab: IGIMFversions}
\end{table*}

The stellar IMF, $\xi_\star(m)$, describes the number of newly formed stars in an embedded star cluster, $dN$, in the mass interval $m$ to $m + dm$. The IMF in embedded star clusters can be mathematically modelled as a three-part power law of the form  
\begin{eqnarray}
    \centering
    \xi_\star(m)=
\begin{cases}
k_{1} m^{-\alpha_{1}} & \mathrm{for} \quad 0.08 \leq m/M_{\odot} < 0.50\, , \\
k_{2} m^{-\alpha_{2}} & \mathrm{for} \quad 0.50 \leq m/M_{\odot} < 1.00\, , \\
k_{3} m^{-\alpha_{3}} & \mathrm{for} \quad 1.00 \leq m/M_{\odot} < m_{\mathrm{max}}\, , \\
\end{cases}
    \label{eq: IMF}
\end{eqnarray}
where $k_{i}$ are normalization constants (ensuring continuity of the function) and $\alpha_{i}$ are the power-law indices or slopes. The canonical IMF \citep{Kroupa_2001} has $\alpha_{1} = 1.3\pm0.3$, $\alpha_{2} = \alpha_{3} = 2.3\pm0.36$ \citep[their equation~4.55]{Kroupa_2013}. The uncertainties of the slope of the IMF can have different interpretations. It may originate from an environment-dependent IMF (\citealt{Kroupa_2013} their section 9.5 and \citealt{Yan_2023}) or a random IMF difference between different star clusters \citep{2022A&A...666A.113D}. For this study, we consider only the average shape of the IMF and the dependence of this shape on the star-formation environment. The values of $\alpha_{1}$ and $\alpha_{2}$ have been verified by \citet{Reid_2002} and \citet{Kirkpatrick_2023} who suggest a possible flattening below $0.22\,M_{\odot}$. The maximum mass of a star in a cluster, $m_{\mathrm{max}}$, is given by the $m_{\mathrm{max}} - M_{\mathrm{ecl}}$ relation \citep[][]{Weidner_2006,Yan_2023}, which states that $m_{\mathrm{max}}$ depends on the cluster stellar mass $M_{\mathrm{ecl}}$.

Observations of the Milky Way showed that about $70-90\%$ of all stars are formed in embedded clusters in giant molecular clouds \citep{Lada_2003}, while the other $10-30\%$ may be stars formed in short-lived clusters \citep{Recchi_2009, Dinnbier_2022}. 
The embedded cluster mass function (ECMF) is described by a single power law \citep[][]{Lada_2003,Lieberz_2017} of the form
\begin{equation}\label{eq: ECMF}
\xi_{\mathrm{ecl}}(M_{\mathrm{ecl}},\psi) =
k_{\mathrm{ecl}} M_{\mathrm{ecl}}^{-\beta(\psi)}~~~~\mathrm{for}~~~~5~M_\odot \leq M_{\mathrm{ecl}} < M_{\mathrm{ecl,max}}(\psi), \\
\end{equation}
where $k_{\mathrm{ecl}}$ is the normalization constant, $5~M_\odot$ is the minimum mass of an embedded cluster \citep[e.g.][]{Joncour_2018} and $M_{\mathrm{ecl,max}}$ is the maximum mass of an embedded star cluster forming a galaxy with SFR $\psi$. The power-law index of the ECMF is about -2 \citep{Lada_2003} with a possible variation in galaxies with different SFRs as suggested in \citet{2011MNRAS.412..979W,2013MNRAS.436.3309W}. We refer the readers to \citet{Yan_2017} for more details. JWST observations will better constrain the ECMF and its possible variation.

Approximating that all stars of the galaxy are formed in embedded star clusters \citep{Kroupa_1995}, 
the IGIMF theory \citep{Kroupa_2003} gives the gwIMF, $\xi_{\rm ecl}$, by adding up the stellar IMFs in all embedded clusters of a galaxy, that is, combining the stellar IMF of each embedded star cluster (Eq.~\ref{eq: IMF}) and the ECMF (Eq.~\ref{eq: ECMF}):
\begin{eqnarray}
    \xi_{\mathrm{gal}}(m, \psi, \rm{Z})~=~ \\ \nonumber
    =\int_{0}^{+\infty} \xi_\star(m, M_{\mathrm{ecl}}, \mathrm{Z})~\xi_{\mathrm{ecl}}(m, M_{\mathrm{ecl}}, \psi) {\rm d} M_{\mathrm{ecl}},
    \label{eq:IGIMF_basic}
\end{eqnarray}
where Z is the metal mass fraction.

We note that the exact dependency of the shape of the IMF and ECMF on the environment is under debate (cf. \citealt{Fontanot_2018,Sneppen_2022}). For example, the original IGIMF version as proposed by \citet{Kroupa_2003} and \citet{Weidner_2006} adopted the invariant IMF for low-mass stars ($\alpha_{1} = 1.3$, $\alpha_{2} = 2.3$) while \citet{Jerabkova_2018} adopt variable IMF slopes with $\alpha_{1}$ and $\alpha_{2}$ coefficients that are functions of the stellar iron abundance in the logarithmic scale. Different from the previous works, \citet{Yan_2020,Yan_2021} assume that $\alpha_{1}$ and $\alpha_{2}$ depends on Z
\begin{equation}\label{eq:alpha_12}
\begin{split}
    &\alpha_1=1.3+\Delta\alpha \cdot (\mathrm{Z}-\mathrm{Z}_{\odot}) \quad  \mathrm{for} \quad  0.08 \leq m/M_{\odot} < 0.50\, ,\\
    &\alpha_2=2.3+\Delta\alpha \cdot (\mathrm{Z}-\mathrm{Z}_{\odot}) \quad  \mathrm{for} \quad  0.50 \leq m/M_{\odot} < 1.00\, ,
\end{split}
\end{equation}
where $\Delta \alpha = 63$ is a constant constrained by observation \citep{Yan_2020,Yan_2021}. $Z$ and
$Z_{\odot} = 0.0142$ \citep{Asplund_2009} are the mean stellar metal-mass fraction of the system and the Sun, respectively.
Although a strong IMF--metallicity correlation is reported \citep{Geha_2013,2015ApJ...806L..31M}, the dependency of the IMF on the stellar metallicity is different in different studies \citep{Yan24}, probably due to systematic errors of the measurement methods or hidden parameters \citep{MartinNavarro_2019}.
The power-law index of the high-mass stars ($1.00 \leq m/M_{\odot} < m_{\mathrm{max}}$) is given by \citep{Marks_2012,Marks_2014}
\begin{eqnarray}
     \centering
     \alpha_{3}=
 \begin{cases}
 2.3  & \mathrm{for} \quad x < -0.87\, , \\
 -0.41 x + 1.94 & \mathrm{for} \quad x \geq -0.87\, , \\
 \end{cases}
     \label{eq:alpha_3}
 \end{eqnarray}
 with
 \begin{eqnarray}
     \centering
     x~=~-0.14 [Z] + 0.6 \log_{10}\bigg(\frac{M_{\mathrm{ecl}}}{10^{6}M_{\odot}} \bigg)+2.82,
     \label{eq:IGIMF_rho_cl_x}
 \end{eqnarray}
where $M_{\mathrm{ecl}}$ is the initial stellar mass of the star cluster and ${\rm [Z]}\equiv{\rm log}_{10}({\rm Z}/{\rm Z}_{\odot})$. The expression of $x$ on $M_{\rm ecl}$ was derived in \citet[their equation 9]{Jerabkova_2018}. Here we have the original IMF shape dependency on iron abundance modified to a dependency on metallicity \citep{Yan_2020} because carbon and oxygen are more abundant and more important metal coolants for the star formation process \citep{2022MNRAS.509.1959S}. A summary of IMF and ECMF variations in different IGIMF models is provided in Table~\ref{tab: IGIMFversions}. 

\begin{figure*}[!hbt]
    \includegraphics[width=\linewidth]{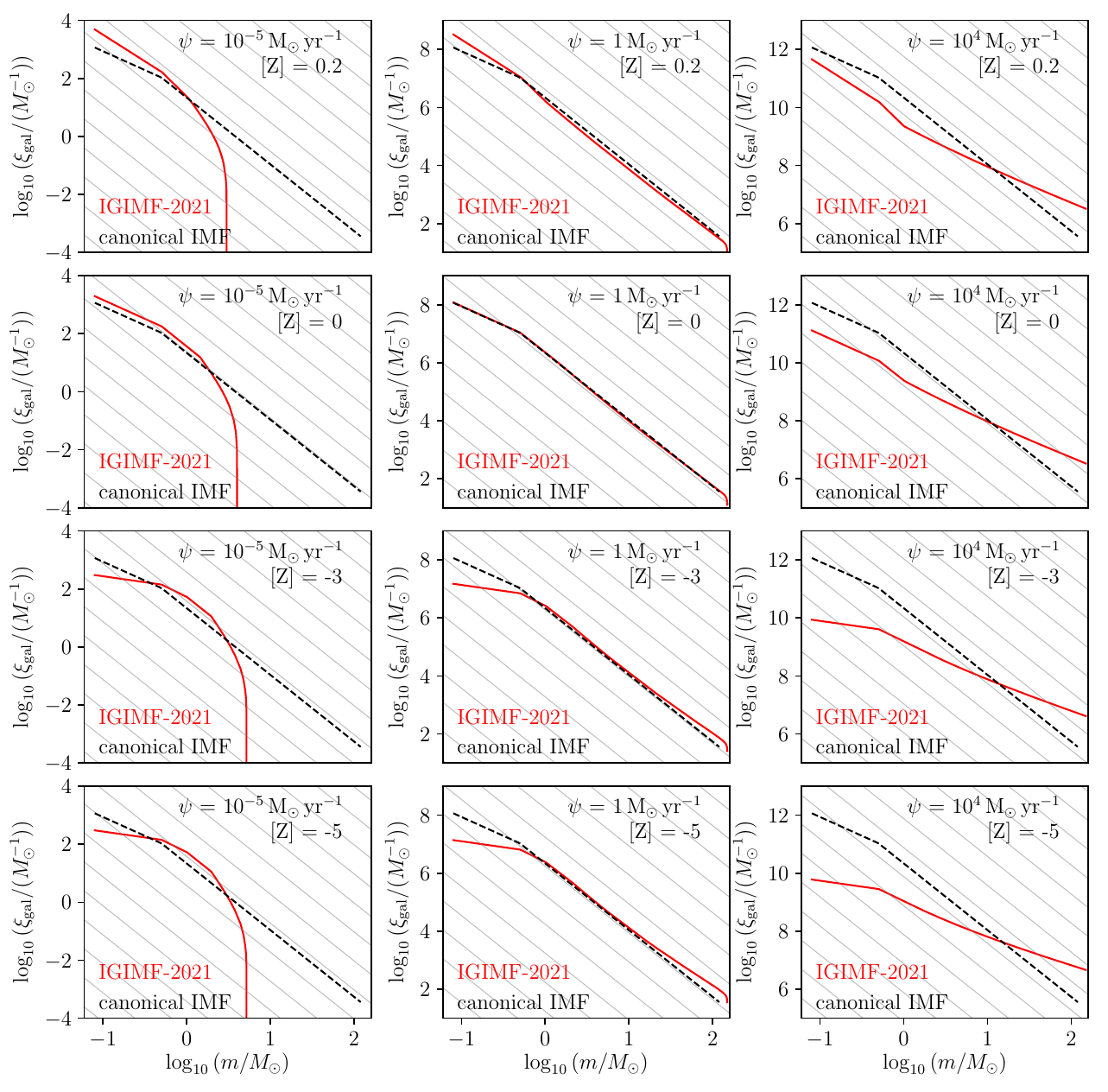}
    \caption{Grid of simulated gwIMFs in dependence of the SFR (from left to right: $\psi = 10^{-5}, 1, 10^{4}~\rm{M_{\odot}\,yr^{-1}}$) and metallicity (from top to bottom: [Z] = 0.2, 0, -3, -5) for the IGIMF-2021 formulation (solid red line) by \citet{Yan_2021} using the GalIMF code \citep{Yan_2017} in comparison with the canonical IMF (dashed black) with the power-law indices $\alpha = 1.3$ for stars in the mass range of $0.08-0.5\,M_{\odot}$ and $\alpha=2.3$ for stars with $> 0.5\,M_{\odot}$ \citep[][]{Kroupa_2001}. The grey solid lines refer to $\alpha=2.3$ for visualization. This grid is similar to Fig.~2 of \citet{Jerabkova_2018} but includes only the IGIMF-2021 formulation which is adopted throughout the analysis presented in this study.}
    \label{figure_gwIMF}
\end{figure*}

For this study, we adopt the most up-to-date IGIMF-2021 model as applied in \citet{Yan_2021}.
The effect of the environment-dependent stellar gwIMF, IGIMF-2021, is qualitatively investigated in Fig.~\ref{figure_gwIMF} by showing a grid of gwIMFs for different global SFRs in the range of $\psi=[10^{-5}, 10^{4}]\,\rm{M_{\odot}\,yr^{-1}}$ and metallicities in the range of $[Z]=[-5, 0.2]$, where $[Z] = \log_{10}(Z/Z_{\odot})$ is the metallicity normalized to the solar value $Z_{\odot} = 0.0142$ \citep{Asplund_2009}. This grid is similar to the analysis of Fig.~2 by \citet{Jerabkova_2018} but includes only the newer IGIMF-2021 formulation \citep{Yan_2021} that results in bottom-heavier IMFs for metal-rich galaxies than IGIMF-2018, and therefore, a higher $M_\star/L$ (discussed in Section~\ref{subsec:Stellar mass-to-light ratio} below). 

Generally, the gwIMF constructed from the IGIMF-2021 (red lines in Fig.~\ref{figure_gwIMF}) systematically varies with the global SFR and metallicity but as it is constructed based on observational constraints, the gwIMF is consistent with the canonical IMF at Solar metallicity and $\psi = 1 \,\rm{M_{\odot}\,yr^{-1}}$ (dashed black line). The gwIMFs of the IGIMF-2021 formulation become bottom-heavy for $[Z] > 0$ and bottom-light for $[Z]<0$ independent of the SFR, reproducing the observed correlation \citep{Geha_2013,Parikh_2018,Zhou_2019,Yan24}. The gwIMF becomes top-heavy for $\psi \ga 10\,\rm{M_{\odot}\,yr^{-1}}$ and top-light for $\psi\la0.1\,\rm{M_{\odot}\,yr^{-1}}$, following the formulation of \citet{Marks_2012}. As illustrated more clearly in Fig.~B1 of \citet{Yan_2017} the top-heaviness of the gwIMFs for $\psi > 1\,\rm{M_{\odot}\,yr^{-1}}$ slightly increases with decreasing metallicity.

We note that equation~(\ref{eq:alpha_12}) is only tested and constrained by observations of stellar populations with $-3<{\rm [Z]}<0.25$. The extrapolated IMF slopes for ${\rm [Z]} > 0.3$ have not been confirmed by observations and may not be correct.

\subsection{Galaxy chemical evolution model} \label{subsubsec:Chemical enrichtment}

The GalIMF code developed by \citet{Yan_2019a,Yan_2021} is a Python 3 package that couples a gwIMF generator (galimf.py) with a chemical evolution model (galevo.py). The codes are introduced and described in full detail in \citet{Yan_2017} and \citet{Yan_2019a}, respectively, and are publicly available on GitHub. We provide here a summary of the most important assumptions and model parameters.  

The chemical evolution calculation assumes a monolithic collapse for galaxy formation whereby primordial gas is transformed into living stars which subsequently evolve into white dwarfs, neutron stars, and black holes over cosmic time. In this study, the gas phase is modelled as a single-zone gas, which is continuously enriched by stellar evolution. 
We also assume that galactic outflows are turned off to reduce the number of free parameters and simplify the discussion. The survey by \citet{Valentino_2021} demonstrates feedback by AGN to have a minimal effect on the gas, dust fractions, and star formation efficiencies suggesting the above assumption to be realistic. Furthermore, studies of star-forming galaxies including dwarfs show that outflows are local and do not affect the evolution significantly: thus, star-bursting dwarfs have similar baryonic and neutral gas fractions to those of typical dwarfs \citep{Lelli_2014}, and 
they also have similar properties of their ionized gas content \citep{Marasco_2023}. Dwarf and massive star-forming galaxies at redshift $z=1-2$ are observed to have negligible large-scale outflows \citep{Concas_2022}. A review of the relevance of feedback in star-forming dwarf galaxies can be found in \citet{Lelli_2022}. Thus supernova feedback has an observable effect in star-forming galaxies, but it's largely local. Supernova feedback stirs the gas inside the potential well and drives non-circular motions in the disk plane. It may also lead to outflows outside the disk plane, but the gas mostly likely falls back on the disk in the galactic fountain fashion, creating a small-scale gas circulation that can help metal mixing and gas recycling. Our neglect of outflows is thus a realistic aspect of the models presented here.

Similar to \citet{Yan_2021}, the main input parameters of the galaxy evolution module are the SFH and the gas-convergence factor, $g_{\rm conv}$, which is the 
ratio between the total stellar mass ever formed and the initial gas mass reservoir that is involved in the instantaneous well-mixing assumption. The $g_{\rm conv}$ value quantifies the SFE and correlates with the metal enrichment of the simulated galaxy. A higher value of $g_{\rm conv}$ increases the metallicity while a lower value decreases the metallicity of the galaxy. Therefore, the adopted $g_{\rm conv}$ value can be constrained by the observed metallicities of galaxies. For example, \citet[their figure 8]{Yan_2021} suggest that $g_{\mathrm{conv}} = 0.25$ for elliptical galaxies.

The timestep of the simulation is set to $\delta t=10$~Myr, which corresponds, approximately, to the time span for large molecular clouds to collapse and form a population of embedded star clusters that can well represent the ECMF \citep{Weidner_2004b,Yan_2017}.
For each 10~Myr time step, the chemical evolution module updates the chemical abundances, the mass of stellar remnants, living stars, and gas mass by accounting for the contributions of stellar populations formed in all previous timesteps. Then the code calculates the instantaneous environment-dependent gwIMF as a function of $\psi$ and Z according to equation~(\ref{eq:IGIMF_basic}) for the stars formed at this timestep.

\subsection{SFHs of the local star-forming galaxies} \label{subsubsec:Star formation histories}

Converting the observed luminosities of galaxies to stellar masses requires the mass-to-light ratio of the stellar population, $M_\star/L$ (including remnants), which mainly depends on the underlying stellar IMF, SFH, and chemical evolution. In the case of an invariant gwIMF, $M_\star/L$ will depend on the shape of the SFH, but not on the absolute SFR values. On the other hand, with a SFR-dependent gwIMF, $M_\star/L$ also depends on the absolute SFR \citep{Zonoozi_2019}. 

The real SFHs of observed galaxies are uncertain. Main sequence galaxies can have rising or declining SFRs. Assuming a rising SFR could result in an estimated present-day SFR that is 0.3 to 0.6 dex higher than the value when assuming a declining SFH \citep[Section~3.1.4 of][]{Speagle_2014}. On the other hand, \citet{Kroupa_2020} showed that most of the star-forming galaxies in the LV have near constant SFRs over a time-span of $12\,\rm{Gyr}$. Therefore, it is reasonable to assume as a first approach SFH models with a constant SFR over time for star-forming galaxies. Assuming a fixed SFR is similar to the assumption that a fixed fraction of gas is converted to stars per time interval because the total gas mass in our closed-box model does not change more than 25\%.
In addition, we assume that all galaxies start forming their stars 0.2~Gyr after the Big Bang till the present time. Thus, throughout the analysis, the present-day properties of galaxies refer to an age of $13.6$~Gyr. 

A more comprehensive modelling of galaxies with different SFHs and ages is needed to extend the current study. Extreme SFH differences may lead to a $M_{\star}/L$ variation by a factor of 2 \citep[fig 5 in][]{2001ApJ...550..212B} which is similar to the variation caused by a variable gwIMF for a constant SFH (Section~\ref{subsec:Stellar mass-to-light ratio}). In addition, the $M_{\star}/L$ value depends on the photometric band. The here-developed photGalIMF code will be able to perform such analysis after the inclusion of more photometric bands with the present study only employing the Ks band to demonstrate the potential mass estimation effect when a variable IMF is considered. The discussion presented here is useful because adopting a constant NIR mass-to-light ratio without taking into account the variation of the IMF and SFH is common in practice \citep{McGaugh_2014}. A systematic overestimation of stellar mass due to an unrealistic SFH would not affect our main conclusions.

\subsection{photGalIMF: Evolution model of the stellar luminosity} \label{subsubsec:stellar evolution}
The previous GalIMF code only considers the zero-age main sequence luminosity \citep[Section~4.4 of][]{Yan_2019a}. To include the metallicity and age dependencies of stellar luminosities for pre-main sequence and giant stars, we develop the photGalIMF code\footnote{\url{https://github.com/juzikong/photGalIMF}\label{link:Github_photgalIMF}} which extends the GalIMF code\footnote{\url{https://github.com/Azeret/galIMF}\label{link:Github_galIMF}} by coupling it with the stellar isochrones of PARSEC \citep{Bressan_2012} and COLIBRI \citep{Marigo_2013}, which will correctly increase the luminosity estimation of a stellar population up to a factor of a few \citep{Girardi_2013}. The concept of the photGalIMF code is illustrated in Fig.~\ref{figure_photGalIMF_flowchart}.

The PARSEC code models the evolutionary tracks from the pre-main sequence phase to the first thermal pulse or carbon ignition. The COLIBRI code adds the thermally-pulsing asymptotic giant branch (TP-AGB) evolution from the 1st thermal pulse to the total loss of envelope. In particular, the current version of the photGalIMF code implements the PARSEC \citep{Bressan_2012} version 1.2s, and the COLIBRI S\_37 code \citep{Pastorelli_2020} for metallicities in the range of $0.008 \leq Z \leq 0.02$, COLIBRI S\_35 \citep{Pastorelli_2019} for $0.0005\leq Z\leq0.006$, COLIBRI PR16 \citep{Marigo_2013}, and the models by \citet{Rosenfield_2016} for $Z\leq0.0002$ and $Z\geq0.03$ (see also the description on the CMD web interface version 3.7\footnote{\label{note1}\url{http://stev.oapd.inaf.it/cgi-bin/cmd}}, hereafter CMD V3.7).

Considering the luminosities of galaxies in the Ks-band of the 2MASS sky survey
\citep{Jarrett_2000, Jarrett_2003} as an example, we downloaded from the CMD V3.7 the PARSEC-COLIBRI stellar isochrone tables for the photometric systems labeled ``2Mass + Spitzer (IRAC + MPIS)" \citep{Cohen_2003,Groenewegen_2006}, which list the Ks-band magnitudes with an effective wavelength of $\lambda_{\mathrm{eff}} = 21620.75$~\AA, a bandwidth of $w_{\mathrm{eff}} = 2714$~\AA, and an attenuation relative to the V-band of $A_{\rm Ks}/A_{\mathrm{V}} = 0.11675$ as given by the CMD V3.7. We apply the default options of the CMD V3.7 which provide the bolometric correction ``YBC+new Vega", a dust composition of 60\% Silicate and 40\% AlOx for M stars and 85\% AMC and 15\% SiC for C stars \citep{Groenewegen_2006}, a total extinction of $A_{\mathrm{V}} = 0.0$~mag, and a long-period variability during the RGB and AGB phase as modelled by \citet{Trabucchi_2021}. In total, we incorporated the isochrones for ten different metallicities, Z = 0.0001, 0.0002, 0.0005, 0.001, 0.004, 0.008, 0.01, 0.02, 0.03, and 0.04. 

The Ks-band magnitudes, $M_{\mathrm{Ks}}$, of the so-downloaded and incorporated isochrone tables are converted into luminosities via 
\begin{eqnarray}
    \frac{L_{\mathrm{Ks}}}{L_{\mathrm{Ks},\odot}} = 10^{0.4 (M_{\mathrm{Ks},\odot} - M_{\mathrm{Ks}})} \, ,
    \label{eq:luminsoity_magnitude}
\end{eqnarray}
where $M_{\mathrm{Ks},\odot} = 3.27$ \citep[table~3 of][]{Willmer_2018} is the absolute magnitude of the Sun in the VEGAMAG system.

We calculate the gwIMF for stars formed within a galaxy evolution timestep of $\delta t = 10$~Myr (Section~\ref{subsubsec:Chemical enrichtment}) by assuming that individual stars are randomly forming at different times in this 10 Myr epoch. Therefore, the GalIMF code calculates the time-averaged luminosity of stars formed with a given initial mass, $m$, metallicity, $Z$, and age, $t$, by
\begin{eqnarray}
    \overline{L}_{\mathrm{Ks}}(m,t,Z)~&=&~\frac{1}{\delta t} \int_{t}^{t+\delta t} L_{\mathrm{Ks}}(m,t',Z)~dt'\\ \nonumber 
    &=&\frac{1}{\delta t}\sum_{i=1}^{N} L_{\mathrm{Ks}}(m,t_{i},Z)~\Delta t \, ,
    \label{eq:luminsoity_time_weights}
\end{eqnarray}
where $\delta t = N \cdot \Delta t$, $t_{i} = t+i \cdot \Delta t$ is the age of the isochrone, and $N$ is the number of used isochrones between the time span $t$ and $t + \delta t$ in the GalIMF code. The time resolution of the isochrones is set to $\Delta t = 0.1$~Myr for stars with ages of $t < 1$~Gyr and $\Delta t = 1$~Myr for $t\geq 1$~Gyr. 

The photGalIMF module assigns each formed star in the GalIMF module a time-averaged luminosity value $\overline{L}_{\mathrm{Ks}}$ according to its $m$, $Z$, and $t$. To reduce the computational cost, we linearly interpolate the stellar-mass-time-averaged-luminosity relation (MLR) for stars with $t < 10^{8.5}$~yr, ensuring that the pre-main sequence is accurately described. For $t \geq 10^{8.5}$~yr, the main sequence phase of the MLR is fitted with a $10$-th order polynomial. The giant phase is still linearly interpolated to ensure that the TP-AGB is accurately traced. The sum of all $\overline{L}_{\mathrm{Ks}}$ values is the total luminosity of the simulated galaxy. We acknowledge that the computational efficiency can be further improved by adopting a spline fit of the MLR instead of linear interpolation.

The Ks-band stellar (including remnants) mass-to-light ratio at a given time is then
\begin{eqnarray}
    \centering
    \frac{M_\star}{L_{\mathrm{Ks}}} = \frac{M_{\star,\mathrm{living}} + M_{\mathrm{rem}}}{L_{\mathrm{Ks}}} \, ,
        \label{eq:ML}
\end{eqnarray}
where $M_{\star,\mathrm{living}}$ and $M_{\mathrm{rem}}$ are the mass of the living stars and stellar remnants (i.e. white dwarfs, neutron stars, black holes, calculated as in \citealt{Yan_2019a}), respectively. 

The current version of the photGalIMF code is freely available on GitHub\footref{link:Github_photgalIMF} and provides the galaxy luminosity evolution in the Ks-band of the 2Mass sky survey \citep{Cohen_2003}, the IRAC [3.6]-band of the Spitzer survey \citep[see e.g.][]{Groenewegen_2006}, and the V-band of the UBVRIJHK photometric system \citep{Bessell_1990, MaizApellaniz_2006}. As an example, Fig.~\ref{figure_evolution_MLR_canonicalIMF} shows the time evolution of $M_\star/L_{\mathrm{Ks}}$ for stellar populations with fixed metallicities (Z = 0.0001, 0.004, 0.02, and 0.04) given by the photGalIMF code. The results are qualitatively in agreement with previous studies (e.g. \citealt{Maraston_2005} and \citealt{2015A&A...575A.128B}). More bands including the \textit{Gaia} and JWST photometries, as well as the hydrogen line emissions, will be added in the future.

\subsection{Calibration of the H$\alpha$ luminosity as a SFR tracer}\label{subsec: SFR-Halpha relation}

\begin{figure*}[!hbt]
    \includegraphics[width=\linewidth]{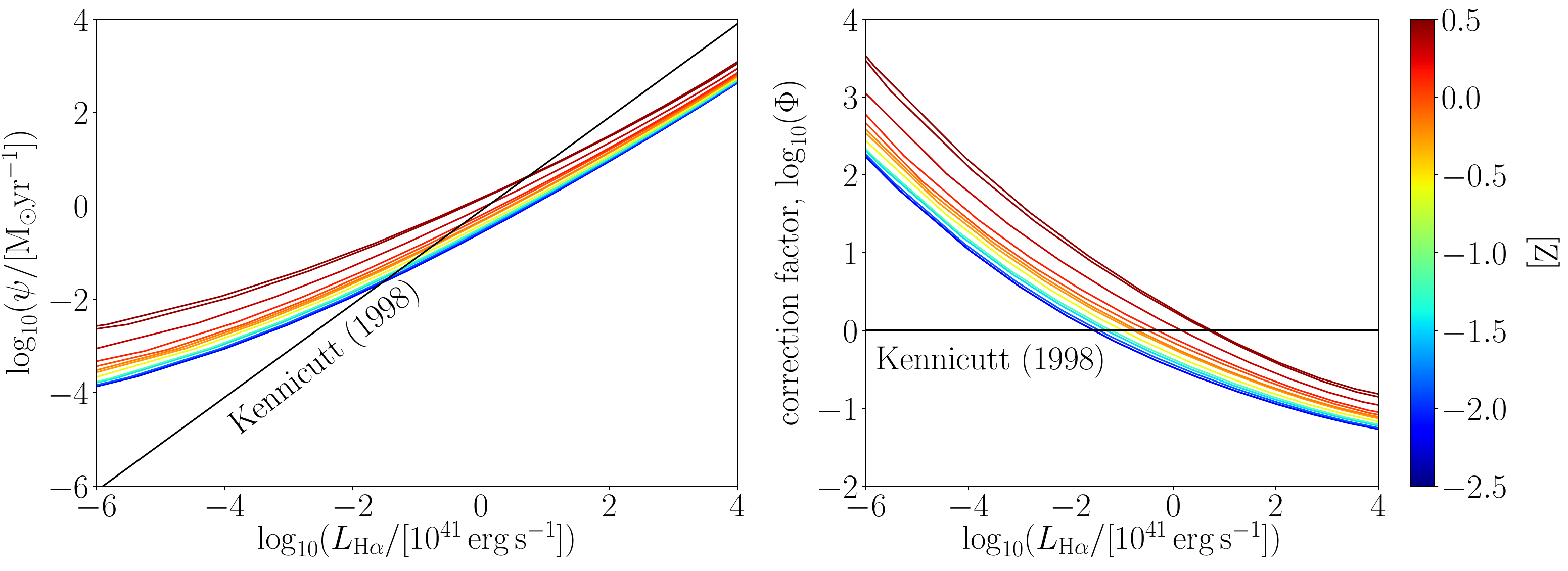}
    \caption{SFR-H$\alpha$ luminosity relations of the canonical invariant IMF and the IGIMF-2021 theory. Left panel: Star formation rate in dependence of the H$\alpha$ luminosity assuming a canonical IMF \citep[solid black line;][]{Kennicutt_1998} and the IGIMF-2021 theory (coloured lines) for different metallicities [Z] shown on the right colorbar. The H$\alpha$ fluxes are calculated with the P\'{E}GASE code (see text). Right panel: Conversion factor of the SFRs, $\Phi = \frac{\psi_{\mathrm{IGIMF}}}{\psi_{\mathrm{K98}}}$, in dependence of the H$\alpha$ luminosity. The solid black line refers to $\psi_{\mathrm{IGIMF}} = \psi_{\mathrm{K98}}$. The fitting parameters are listed in Table~\ref{tab:fitting_parameters}.}
    \label{figure_main_sequence_IGIMF}
\end{figure*}

Equation~(2) of \citet{Kennicutt_1998} empirically relates the SFR and the H$\alpha$ nebular gas luminosity of a galaxy via
\begin{eqnarray}
    \centering
    \frac{\psi_{\mathrm{K98}}}{\rm{M_{\odot} \, yr^{-1}}}~=~7.9 \times 10^{-42} \frac{L(\rm{H\alpha})}{\rm{erg \, s^{-1}}} \, .
    \label{eq:Kennicutt}
\end{eqnarray}
This relation assumes the solar abundance and the Salpeter IMF \citep{Salpeter_1955} and is not valid for a different gwIMF or metallicity \citep{PflammAltenburg_2007, PflammAltenburg_2009a, Jerabkova_2018}.

To estimate the galactic H${\mathrm{\alpha}}$ luminosity, we apply a separate stellar population synthesis code because the current version of the photGalIMF code does not include the calculation of nebular gas H${\mathrm{\alpha}}$ luminosity. 
Following Section~4.3 of \citet{Jerabkova_2018}, we use the second version of the Programme d’\'{E}tude des GAlaxies par Synth\`{e}se \'{E}volutive (P\'{E}GASE) stellar population synthesis code \citep{Fioc_PEGASE.2_1999, Fioc_2011} in combination with the PyPegase python wrapper\footnote{\url{https://github.com/coljac/pypegase} (downloaded on 03.06.2022)} to compute the SFR--H$\alpha$-luminosity relation for the here applied IGIMF-2021 formulation. 
The computation of this conversion uses the gwIMF as an input and calculates an average nebular gas H$\alpha$ flux of a galaxy from the ionizing photons, assuming a constant SFR. Since the code only allows the implementation of a multi-power-law IMF, we adopted the same procedure developed in \citet{Jerabkova_2018} where they approximated the gwIMF as a multi-part power law by fitting the power-law indices in the mass ranges of $0.08 - 0.5 \, M_{\odot}$, $0.5 -1.0 \, M_{\odot}$, $1.0 - 0.8~m_{\mathrm{max}}  \, M_{\odot}$, and $0.8~m_{\mathrm{max}} - 150\,M_{\odot}$ (see their section~4.3). The code calculates the average H$\alpha$ flux value for a timescale over 100~Myr, as short-lived massive stars contribute most significantly to the ionizing flux.

The SFR--H$\alpha$-luminosity relations of the IGIMF-2021 formulation for different metallicities are presented in the left panel of Fig.~\ref{figure_main_sequence_IGIMF}. In general, galaxies with $L_{\mathrm{H\alpha}} \gtrsim 10^{41}$~ergs/s ($L_{\mathrm{H\alpha}} \lesssim 10^{41}$~ergs/s) have lower (higher) SFRs compared to the SFRs derived from \citet{Kennicutt_1998}. This is consistent with figure~7 of \citet{Jerabkova_2018}. 

Following \citet{Kennicutt_1998}, our provided relations based on the continuous-star-formation approximation only consider a systematic gwIMF variation and metallicity differences of the galaxies. Dust content, SFH, and other galaxy properties are assumed to be identical or irrelevant. Therefore, the SFR--H$\alpha$-luminosity relation can be different, for example, if massive galaxies have stronger dust attenuation \citep{2016ApJ...817L...9N}. Essentially, the photometry provided by the photGalIMF code assumes the effect of dust has been removed \citep[see chapters in][]{Zezas_2021}.

To convert the observed H$\alpha$ luminosities to SFRs within the IGIMF-2021 context, we provide, in the following, different fitting functions. 
The SFR--H$\alpha$-luminosity relations for different metallicities are fitted in the range of $-6 < \log_{10}\big(L_{\mathrm{H\alpha}}/(10^{41}\,\rm{erg \,s^{-1}})\big) < 4$ with a fifth-order polynomial in $\log_{10}$-space of the form
\begin{eqnarray}
    \centering
    \log_{10}\bigg(\frac{\psi_{\mathrm{IGIMF}}}{\rm{M_{\odot} \, yr^{-1}}} \bigg)~=~a_{\psi} x^5 + b_{\psi} x^4 + c_{\psi} x^3 +d_{\psi} x^2 + e_{\psi} x + f_{\psi}  \, ,
    \label{eq:IGIMF_Halpha_SFR_fit}
\end{eqnarray}
where $x \equiv \log_{10}\big(L_{\mathrm{H\alpha}}/(10^{41}\,\rm{erg \,s^{-1}})\big)$. The conversion factor between the SFRs based on the linear Kennicutt law \citep{Kennicutt_1998} and the IGIMF theory is defined as \citep[equation~(17) of][]{Jerabkova_2018}
\begin{eqnarray}
    \centering   \Phi(L_{\mathrm{H\alpha}})~=~\frac{\psi_{\mathrm{IGIMF}}}{\psi_{\mathrm{K98}}} \, .
    \label{eq:IGIMF_SFR_correction}
\end{eqnarray}
and its $L_{\mathrm{H\alpha}}$ dependence is shown in the right panel of Fig.~\ref{figure_main_sequence_IGIMF}. This relation is fitted also in the range of $-6 < \log_{10}\big(L_{\mathrm{H\alpha}}/(10^{41}\,\rm{erg \,s^{-1}})\big) < 4$ with a fifth-order polynomial
\begin{eqnarray}
    \centering
 \log_{10}\bigg(\Phi(L_{\mathrm{H\alpha}}) \bigg)~=~a_{\phi} y^5 + b_{\phi} y^4 + c_{\phi} y^3 +d_{\phi} y^2 + e_{\phi} y + f_{\phi}  \, ,
    \label{eq:IGIMF_Kennicutt_SFR_fit}
\end{eqnarray}
where $y\equiv\log_{10}\big(L_{\mathrm{H\alpha}}/(10^{41}\,\rm{erg \,s^{-1}})\big)$.

Finally, the relation between $\psi_{\mathrm{K98}}$ \citep{Kennicutt_1998} and $\psi_{\mathrm{IGIMF}}$ is fitted in the range of $-6 < \log_{10}\big(\psi_{\mathrm{K98}}/(\mathrm{M_{\odot}\,yr^{-1}})\big) < 4$ with a fifth-order polynomial in the $\log_{10}$-space of the form
\begin{eqnarray}
    \centering
    \log_{10}\bigg(\frac{\psi_{\mathrm{IGIMF}}}{\rm{M_{\odot} \, yr^{-1}}} \bigg)~=~a_{\theta} z^5 + b_{\theta} z^4 + c_{\theta} z^3 +d_{\theta} z^2 + e_{\theta} z + f_{\theta}  \, ,
    \label{eq:IGIMF_Kennicutt_SFR_fit_2}
\end{eqnarray}
where $z\equiv\log_{10}\big(\psi_{\mathrm{K98}}/(\mathrm{M_{\odot}\,yr^{-1}})\big)$. 
For example, a galaxy traditionally thought to have $\psi_{\mathrm{K98}}=10^{-5}\,\mathrm{M_{\odot}\,yr^{-1}}$ would have $\psi_{\mathrm{IGIMF}} \approx10^{-3}\,\mathrm{M_{\odot}\,yr^{-1}}$ for $[Z]\approx-0.5$. That is, the SFR based on the nebular ionizing flux may be significantly underestimated for low-SFR galaxies because of a top-light gwIMF (e.g. \citealt{Yan_2017}, their figure 6). Multi-tracer studies of the SFR of dwarf galaxies support the gwIMF predicted by the IGIMF theory \citep{Lee_2009}, as already also emphasized by \citet{PflammAltenburg_2007, PflammAltenburg_2009a} based on the IGIMF-2003 (Table~\ref{tab: IGIMFversions}).

The polynomial coefficients of the fitting functions given by the equations~(\ref{eq:IGIMF_Halpha_SFR_fit}), (\ref{eq:IGIMF_Kennicutt_SFR_fit}), and~(\ref{eq:IGIMF_Kennicutt_SFR_fit_2}) are listed for different metallicities in Table~\ref{tab:fitting_parameters}.

\section{Observations}\label{sec:observations}

\subsection{SFE constrained by the mass--metallicity relation} \label{subsubsec:Mass--metallicity relation}

The gwIMF of a galaxy and the luminosity of stars depend on the metallicity. Therefore, the $M_\star/L$ value of a galaxy depends on the metallicity distribution of the stars calculated by the galaxy chemical enrichment model and the galaxy models need to reproduce the observed mass--metallicity relation of the sample galaxies. 

\begin{figure}[!hbt]
    \includegraphics[width=\columnwidth]{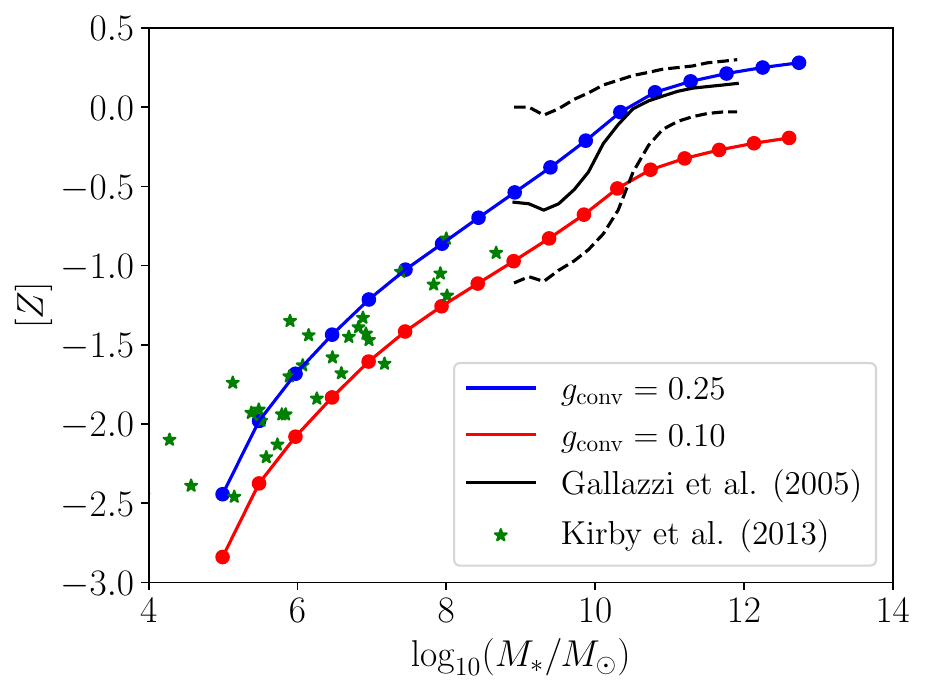}
    \caption{
    Present-day stellar (including remnants) mass--metallicity relation for galaxies with different constant SFRs over a time-span of 13.6~Gyr, adopting $g_{\mathrm{conv}} = 0.10$ (red line) and $g_{\mathrm{conv}}=0.25$ (blue line), within the IGIMF-2021 context. The green stars refer to dwarf galaxies of the Milky Way and M31 galaxies \citep[table~4 of][]{Kirby_2013}. The solid black line shows the median while the dashed lines show the 16th and 84th percentiles of the metallicity distribution of galaxies in the Sloan Digital Sky Survey Data Release 2 \citep[table~1 of][]{Gallazzi_2005}. Note that the shown stellar masses of \citet{Kirby_2013} and \citet{Gallazzi_2005} are derived for the canonical gwIMF but the simulated mass--metallicity relations (red and blue lines) refers to a varying gwIMF within the IGIMF theory (see text).}
    \label{figure_IGIMF_MZ}
\end{figure}

Assuming the closed-box GCE model and constant SFRs (reasons given in Section~\ref{subsubsec:Star formation histories}), between $\psi = 10^{-5}\,\rm{M_{\odot}\,yr^{-1}}$ and $\psi = 10^{3}\,\rm{M_{\odot}\,yr^{-1}}$ in steps of $0.5$~dex, over a time-span of 13.6~Gyr, we calculate for each SFH two different chemical enrichment models with $g_{\mathrm{conv}} = 0.10$ and $g_{\mathrm{conv}} = 0.25$ (as explained in Section~\ref{subsubsec:Chemical enrichtment}).
We find that the observed stellar mass--metallicity relation of star-forming galaxies \citep{Gallazzi_2005,Kirby_2013} lies approximately in between these models as is shown in Fig.~\ref{figure_IGIMF_MZ}. Therefore, we consider that within the framework of our galaxy evolution model, most local star-forming galaxies should have a $g_{\mathrm{conv}}$ value between 0.1 and 0.25. The uncertainties introduced by the unknown $g_{\mathrm{conv}}$ value of galaxies are estimated by applying 0.1 and 0.25 as a lower and upper limit, respectively.

The galactic stellar masses estimated by \citet{Gallazzi_2005} and \citet{Kirby_2013} assume a canonical IMF while our calculations are based on the more realistic empirical environment-dependent gwIMF. However, the calculated masses differ by usually less than a factor of two (Section~\ref{subsec:M/L--L relation}, cf. \citealt{Yan_2021}), which is insignificant compared to the observed scatter in the mass--metallicity relation (Fig.~\ref{figure_IGIMF_MZ}) and, therefore, do not affect our conclusions. Note that the observed mass-metallicity relation is reproduced by the IGIMF-2021 model without invoking outflows. This was already noted by \citet{Koeppen_2007} and is consistent with the observational evidence of starforming and starbursting galaxies lacking large-scale gas outflows even despite the presence of AGN activity (Section~\ref{subsubsec:Chemical enrichtment}).

\subsection{Sample galaxies} \label{subsec: Data}

The observational data are taken from the updated version \citep{Karachentsev_2013} of the Catalogue of Neighbouring Galaxies\footnote{\url{https://www.sao.ru/lv/lvgdb/introduction.php}. Here, we use the latest update from 12.04.2023.} \citep{Karachentsev_2004}, which lists galaxies in the Local Cosmological Volume (LV) defined by Galactocentric distances of $D<11\,\rm{Mpc}$ or radial velocities of $V<600\,\rm{km\,s^{-1}}$. In order to analyse the main sequence of star-forming galaxies (Section~\ref{subsec:main sequence of star-forming galaxies}) and the characteristic stellar mass buildup timescales (Section~\ref{subsec:Stellar-mass buildup times}), we extract the Ks-band luminosity, $L_{\mathrm{Ks}}$, and the SFR based on the integrated H$\alpha$ luminosity assuming the Kennicutt law, $\psi_{\mathrm{K98}}$ \citep{Karachentsev_2013_SFRproperties}. Galaxies marked with $\psi_{\mathrm{K98}}$ limit flags are excluded as only reliable SFR measurements should be included in the analysis. This gives a final sample of 603 galaxies. For calculating the gas depletion timescales (Section~\ref{subsec:Gas depletion timescale}), we also select the measurements of the hydrogen mass, $M_{\mathrm{HI}}$, which reduces the sample to 544 galaxies. The total neutral gas mass of a galaxy is obtained via 
\begin{eqnarray}
    \centering
    M_{\mathrm{gas}}=1.85 M_{\mathrm{HI}},
    \label{eq:Mgas}
\end{eqnarray}
where the prefactor 1.85 accounts for the helium mass and molecular gas mass (cf. section~7 of \citealt{Karachentsev_2013_SFRproperties} and section 2.3.4 of \citealt{2004ApJ...616..643F}).

\section{Results}
\label{sec: Results}

In this section, we calculate the $M_\star/L_{\mathrm{Ks}}$ values and compare the stellar-mass--SFR relation, characteristic stellar mass buildup timescale, and gas depletion timescale of local star-forming galaxies assuming two different frameworks, the invariant canonical IMF and the IGIMF-2021 (Section~\ref{subsec:IGIMF theory}) formulations.

\subsection{Evolution of the mass-to-light ratio of the stellar population} \label{subsec:Stellar mass-to-light ratio}

\begin{figure*}[!hbt]
    \includegraphics[width=\linewidth]{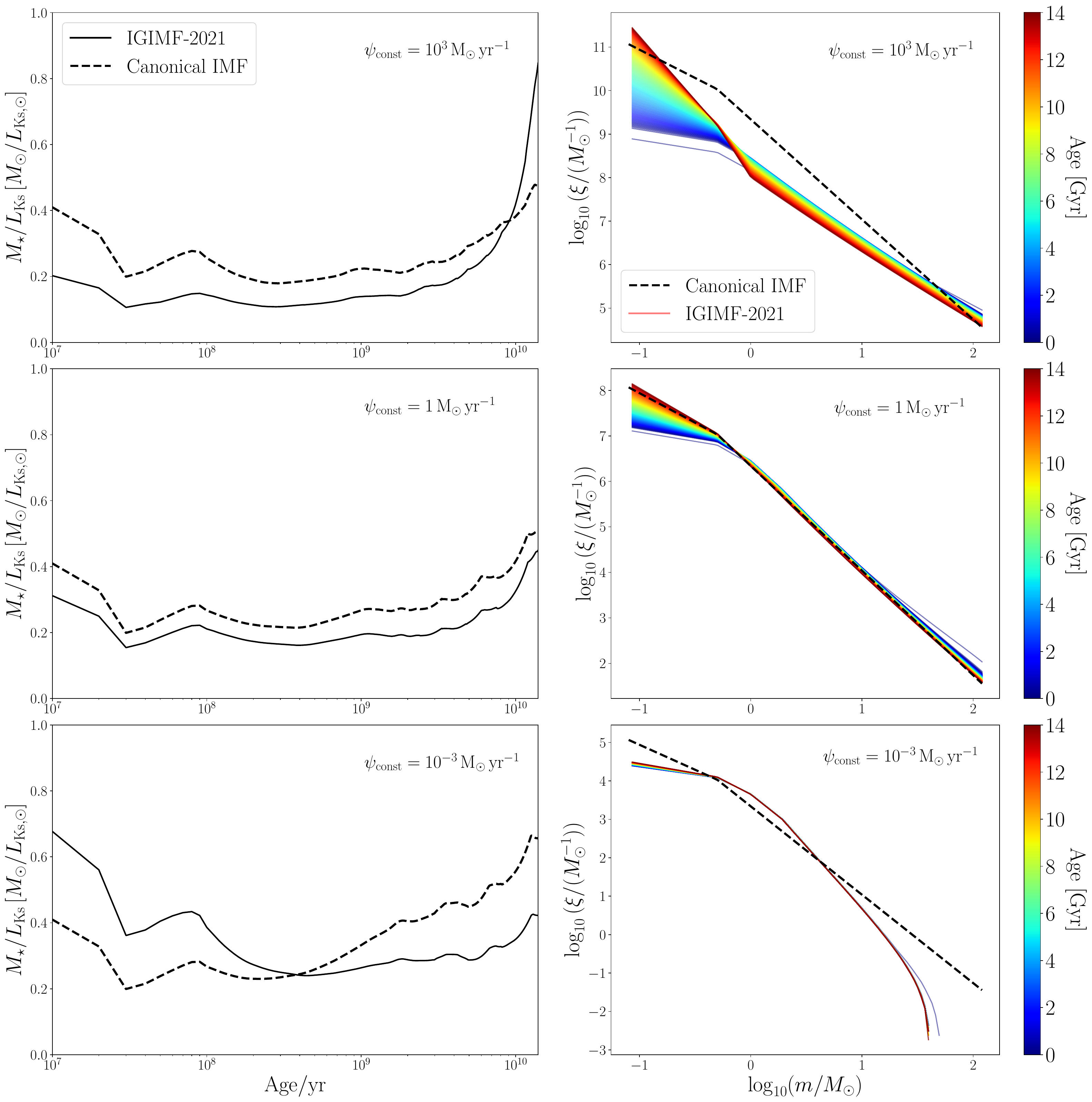}
\caption{Time evolution of the gwIMF (right panels) and the stellar mass-to-light ratio in the Ks-band (left panels) assuming a constant SFR of $\psi = 10^{3}\,\rm{M_{\odot}\,yr^{-1}}$ (top panels), $1\,\rm{M_{\odot}\,yr^{-1}}$ (middle panels), and $10^{-3}\,\rm{M_{\odot}\,yr^{-1}}$ (bottom panels) over a timescale of $14$~Gyr in the invariant canonical gwIMF (dashed black lines in the left and right panels) and IGIMF-2021 (solid black line in the left panels and coloured lines in the right panels) framework. The chemical evolution model adopts $g_{\mathrm{conv}} = 0.25$. The mass-to-light evolution curves adopting the invariant canonical IMF are not identical because they represent the evolution of galaxies with different metal enrichment histories.}
\label{figure_galaxy_evolution_constantSFH_IGIMF}
\end{figure*}

As an example, Figure~\ref{figure_galaxy_evolution_constantSFH_IGIMF} shows the time evolution of the gwIMF and the Ks-band mass-to-light ratio, $M_\star/L_{\mathrm{Ks}}$ (Eq.~\ref{eq:ML}), assuming constant SFRs and $g_{\mathrm{conv}} = 0.25$, constructed assuming an invariant canonical IMF and the IGIMF. The gwIMF of the low-mass stars depends on the stellar metallicity according to the empirical IGIMF-2021 formulation. Compared to the invariant canonical gwIMF case, the IGIMF-2021 model leads to a significant increase of the $M_\star/L_{\mathrm{Ks}}$ value for massive metal-rich galaxies and a lower $M_\star/L_{\mathrm{Ks}}$ value for the low-SFR metal-poor galaxies.

For the IGIMF-2021 formulation and assuming a constant SFR (Section~\ref{subsubsec:Star formation histories}), the $M_\star/L_{\mathrm{Ks}}$ value slowly increases with time after about $2\cdot10^8$~yr, especially for bottom-heavier gwIMFs. We note that the significant increase of $M_\star/L_{\mathrm{Ks}}$ for high-SFR galaxies (top panel in Fig.~\ref{figure_galaxy_evolution_constantSFH_IGIMF}) near 10~Gyr is mainly driven by metal enrichment and the gwIMF dependency on metallicity. 
If a galaxy with a known metallicity has a shorter SFH duration than the 13.6~Gyr assumed above, the time variation of $M_\star/L_{\mathrm{Ks}}$ could lead to an overestimation of the galaxy's stellar mass, by a factor of $\lesssim 3$.

\subsection{M/L--L relation} \label{subsec:M/L--L relation}

\begin{figure*}[!hbt]
\includegraphics[width=\linewidth]{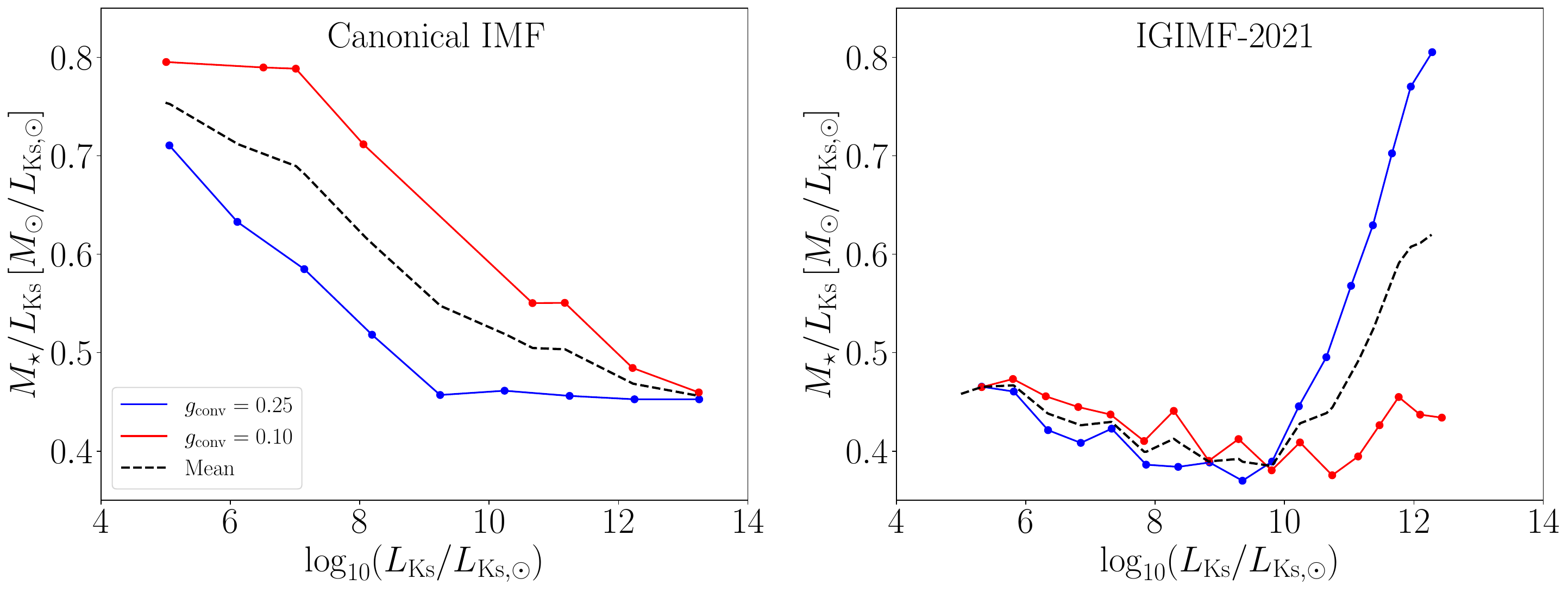}
    \caption{Stellar mass-to-light ratio, $M_\star/L_{\mathrm{Ks}}$ (Eq.~\ref{eq:ML}), in the Ks-band in dependence of $L_{\mathrm{Ks}}$ for galaxies with an age of $13.6$~Gyr assuming $g_{\mathrm{conv}}=25\%$ (red lines) and $10\%$ (blue lines) in the canonical IMF (left panel) and IGIMF-2021 (right panel) context. The dashed black line shows the mean $M_\star/L_{\mathrm{Ks}}$ of the two different chemical enrichment models (red and blue lines).} 
    \label{figure_presentday_ML}
\end{figure*}

The observed Ks-band luminosities listed in the Catalogue of Neighbouring Galaxies (Section~\ref{subsec: Data}) are converted to stellar masses in the IGIMF context by applying the present-day $M_\star/L_{\mathrm{Ks}}$ values shown in the right panel of Fig.~\ref{figure_presentday_ML}. The $M_\star/L_{\mathrm{Ks}}$ nominal-face-value for a given Ks-band luminosity is taken to be the mean value of the chemical enrichment models with $g_{\mathrm{conv}} = 0.10$ and
$g_{\mathrm{conv}} = 0.25$ (Section~\ref{subsubsec:Mass--metallicity relation}). The upper and lower limit of $M_\star/L_{\mathrm{Ks}}$ refer either to the $g_{\mathrm{conv}} = 0.10$ or
$g_{\mathrm{conv}} = 0.25$ model depending on which yields the higher or lower $M_\star/L_{\mathrm{Ks}}$ value. The IMF is affected by the metallicity more significantly in the metal-rich regime. As a consequence, the errorbar of $M_\star/L_{\mathrm{Ks}}$ increases with higher $L_{\mathrm{Ks}}$. Here, the resulting uncertainty denotes the possible range of $M_\star/L_{\mathrm{Ks}}$ for a constant SFR (Section~\ref{subsubsec:Star formation histories}), and if $g_{\mathrm{conv}}$ is between 0.1 and 0.25. For any specific galaxy with an accurate metallicity measurement, its $M_{\star}/L_{\mathrm{Ks}}$ can be estimated with a smaller uncertainty.
Instead of applying the $M_\star/L_{\mathrm{Ks}}$ values in the left panel of Fig.~\ref{figure_presentday_ML}, we adopt a fixed $M_\star/L_{\mathrm{Ks}} = 0.6$ for the canonical IMF \citep{McGaugh_2014} in Sections~\ref{subsec:main sequence of star-forming galaxies} and \ref{subsec:Stellar-mass buildup times}.

The SFRs are IGIMF corrected by using the fitting function defined by equation~(\ref{eq:IGIMF_Kennicutt_SFR_fit_2}). The metallicities of the observed galaxies have not been estimated in the current version of the Catalogue of Neighbouring galaxies. Therefore, we apply for the nominal-face-value of the corrected SFRs equation~(\ref{eq:IGIMF_Kennicutt_SFR_fit_2}) for a Solar metallicity (Table~\ref{tab:fitting_parameters}). The corresponding errorbars are conservatively estimated by adopting the lower and upper limits [Z]$ = -2.20$ and [Z]$= 0.45$, respectively.

The present-day $M_\star/L_{\mathrm{Ks}}$ values of our grid galaxies with two different $g_{\mathrm{conv}}$ parameters are calculated. This corresponds to the $g_{\mathrm{conv}}$ value of the final time step in Fig.~\ref{figure_galaxy_evolution_constantSFH_IGIMF}.
When assuming the invariant canonical gwIMF, the more massive galaxies have smaller $g_{\mathrm{conv}}$ values because they have a higher metallicity (Fig.~\ref{figure_IGIMF_MZ}) that increases the Ks-band luminosity. 
In the IGIMF context, the gwIMF becomes bottom-heavy for metal-rich galaxies. Therefore, the present-day $M_\star/L_{\mathrm{Ks}}$ also increases for the most massive galaxies as demonstrated in Fig.~\ref{figure_presentday_ML}. This gwIMF-metallicity dependency is absent for models assuming $g_{\mathrm{conv}} = 0.1$ that result in metal-poor galaxies because the IMF of low-mass stars is no longer sensitive to metallicity when the metallicity is low enough. Our calculation, considering galaxies with a constant SFR over 13.6~Gyr and an empirical gwIMF variation, suggests that the stellar mass of a galaxy is overestimated by a factor of $\lesssim 2$ for dwarf galaxies, and the galactic stellar mass is underestimated by a factor of $\approx 2$ for the most massive galaxies with a super-Solar metallicity (${\rm [Z]} \approx 0.2$, see Fig.~\ref{figure_IGIMF_MZ}).

\subsection{Stellar-mass--SFR relation of star-forming galaxies} \label{subsec:main sequence of star-forming galaxies}
\begin{figure*}[!hbt]
    \includegraphics[width=\linewidth]{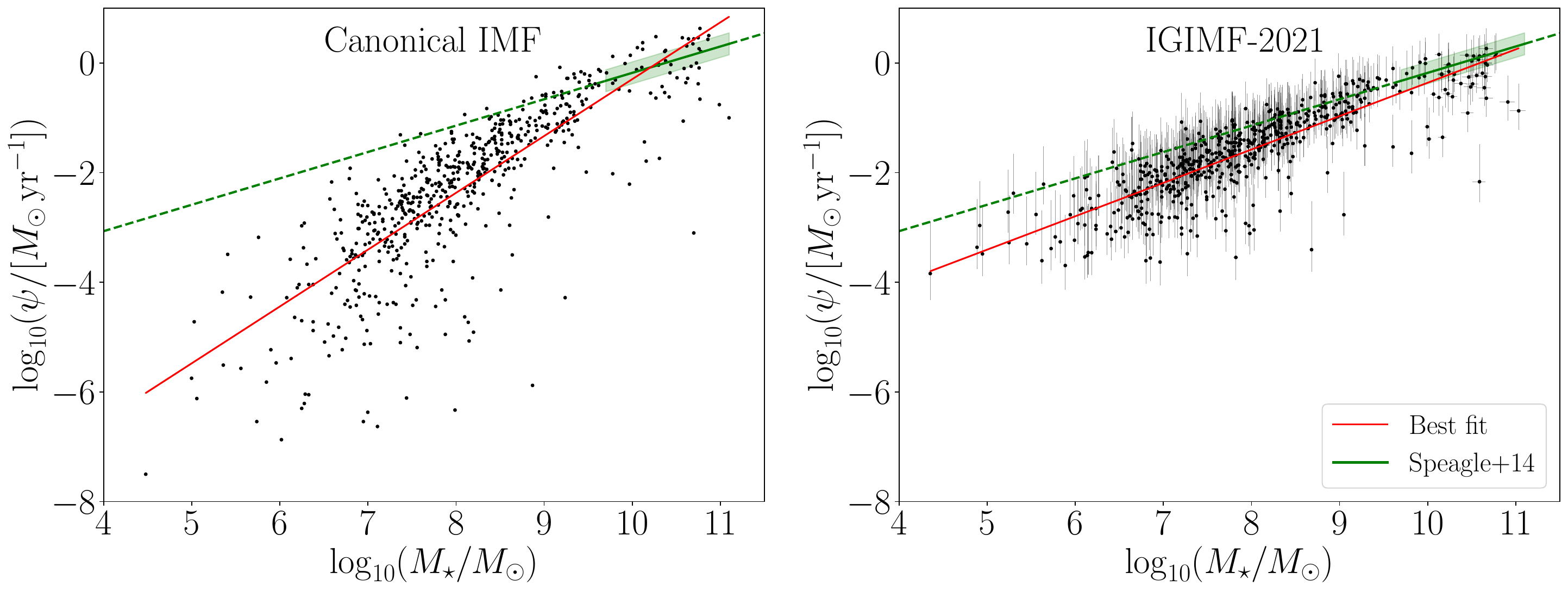}
    \caption{Star formation rate derived from the H$\alpha$ luminosity in dependence of the stellar mass derived from the Ks-band luminosity \citep{Karachentsev_2013} assuming an invariant canonical gwIMF (left panels) and the IGIMF-2021 (right panels) for galaxies located in the LV. The errorbars of the IGIMF-corrected stellar mass values refer to the two different applied chemical enrichment models (Section~\ref{subsec:M/L--L relation}) and the errorbars of the IGIMF-corrected SFR values refer to metallicities of $[Z]=-2.20$ (lower limit) and $[Z]=0.45$ (upper limit; see Section~\ref{subsec:M/L--L relation}). The SFR values have an uncertainty of about $30\%$ due to distance uncertainties \citep[see Fig.~1 of][]{Kroupa_2020,Karachentsev_2004, Karachentsev_2013}. The solid green lines show the present-day main sequence of star-forming galaxies with $10^{9.7}< M_\star/M_{\odot} < 10^{11.1}$ and a scatter of $\pm0.2$~dex highlighted by the shaded area \citep[i.e. equation~(28) of][and equation~(\ref{eq:mainsequencefit_canimf})]{Speagle_2014}. The dashed green lines are extrapolations beyond the adopted fitted range of the stellar masses. The solid red lines show the linear best fit to the corresponding data (Eq.~\ref{eq:mainsequencefit_igimf} for the right panel). Note the smaller scatter in the right panel.}     \label{figure_main_sequence_IGIMF_observations}
\end{figure*}

The stellar-mass--SFR relation of star-forming galaxies in the Local Cosmological Volume assuming the invariant canonical gwIMF and the IGIMF-2021 formulations is presented in Fig.~\ref{figure_main_sequence_IGIMF_observations} by plotting the present-day star formation rates in dependence of the stellar masses derived from the Ks-band luminosities.

In the canonical gwIMF context (left panel of Figure~\ref{figure_main_sequence_IGIMF_observations}), we assume a constant $M_\star/L_{\mathrm{Ks}}=0.6$. The resulting stellar-mass--SFR relation has a large scatter, especially for low-mass galaxies, such that the star-forming main sequence of low-mass galaxies is ill-defined.
Star-forming galaxies with $\psi \ga 10^{-3}\,\rm{M_{\odot}\,yr^{-1}}$ and $M_\star \la 10^{10}\,M_{\odot}$ are consistent with a constant SFR over a star-forming timescale of $\approx12$~Gyr (because $M_{\star} = \psi~\times~12$~Gyr), while more massive galaxies typically follow the main sequence of galaxies \citep{McGaugh_2017,Schombert_2019,Kroupa_2020,Haslbauer_2023a}. For more massive galaxies, \citet{Speagle_2014} quantified the main sequence for a given age, $t$, and found for star-forming galaxies with $10^{9.7} < M_\star/M_{\odot} < 10^{11.1}$ a linear relation of the form (see their equation~(28)),
\begin{eqnarray}
    \centering
    \log_{10}\bigg(\frac{\psi}{\rm{M_{\odot} \, yr^{-1}}} \bigg)~&=&~(0.84 - 0.026\times t)\times\log_{10}\bigg(\frac{M_\star}{M_{\odot}} \bigg) \, , \\ \nonumber 
    &-& (6.51 - 0.11\times t) \, .
    \label{eq:mainsequencefit_speagle}
\end{eqnarray}
For $t = \tau_{\mathrm{h}}$, this relation reduces to
\begin{eqnarray}
    \centering
    \log_{10}\bigg(\frac{\psi}{\rm{M_{\odot} \, yr^{-1}}} \bigg)~=~0.48~\log_{10}\bigg(\frac{M_\star}{M_{\odot}} \bigg)  - 4.99 \, ,
    \label{eq:mainsequencefit_canimf}
\end{eqnarray}
shown by the green dashed line in Fig.~\ref{figure_main_sequence_IGIMF_observations} (cf. \citealt{2020ARA&A..58..661F}).

Assuming the gwIMF is given by the IGIMF-2021 theory, the observed Ks-band luminosities listed in the Catalogue of Neighbouring Galaxies (Section~\ref{subsec: Data}) are converted to stellar masses by applying the present-day $M_\star/L_{\mathrm{Ks}}$ values shown in the right panel of Fig.~\ref{figure_presentday_ML}.
Galaxies with $L_{\mathrm{Ks}} \la 10^{9} \, L_{\mathrm{Ks},\odot}$ ($L_{\mathrm{Ks}} \ga 10^{9} \, L_{\mathrm{Ks},\odot}$) have higher (lower) SFRs compared to a canonical gwIMF as discussed in Fig.~\ref{figure_main_sequence_IGIMF}. 
A tight stellar-mass--SFR relation similar to the main sequence of massive galaxies given in \citet{Speagle_2014} is recovered, extending to low-mass galaxies. Approximating the resulting main sequence by a linear function over $10^{4.4}< M_\star/M_{\odot}<10^{11.1}$ yields,
\begin{eqnarray}
    \centering
    \log_{10}\bigg(\frac{\psi}{\rm{M_{\odot} \, yr^{-1}}} \bigg)~=~0.61~\log_{10}\bigg(\frac{M_\star}{M_{\odot}} \bigg)  - 6.45 \, ,
    \label{eq:mainsequencefit_igimf}
\end{eqnarray}
as is shown by the red solid line in the right panel of Fig.~\ref{figure_main_sequence_IGIMF_observations} (see also Fig.~1.8 of \citet{Kroupa_Jerabkova_2021} demonstrating the IGIMF correction for the galaxy main sequence at different redshifts).

\subsection{Gas depletion timescale} \label{subsec:Gas depletion timescale}
\begin{figure*}[!hbt]
    \includegraphics[width=\linewidth]{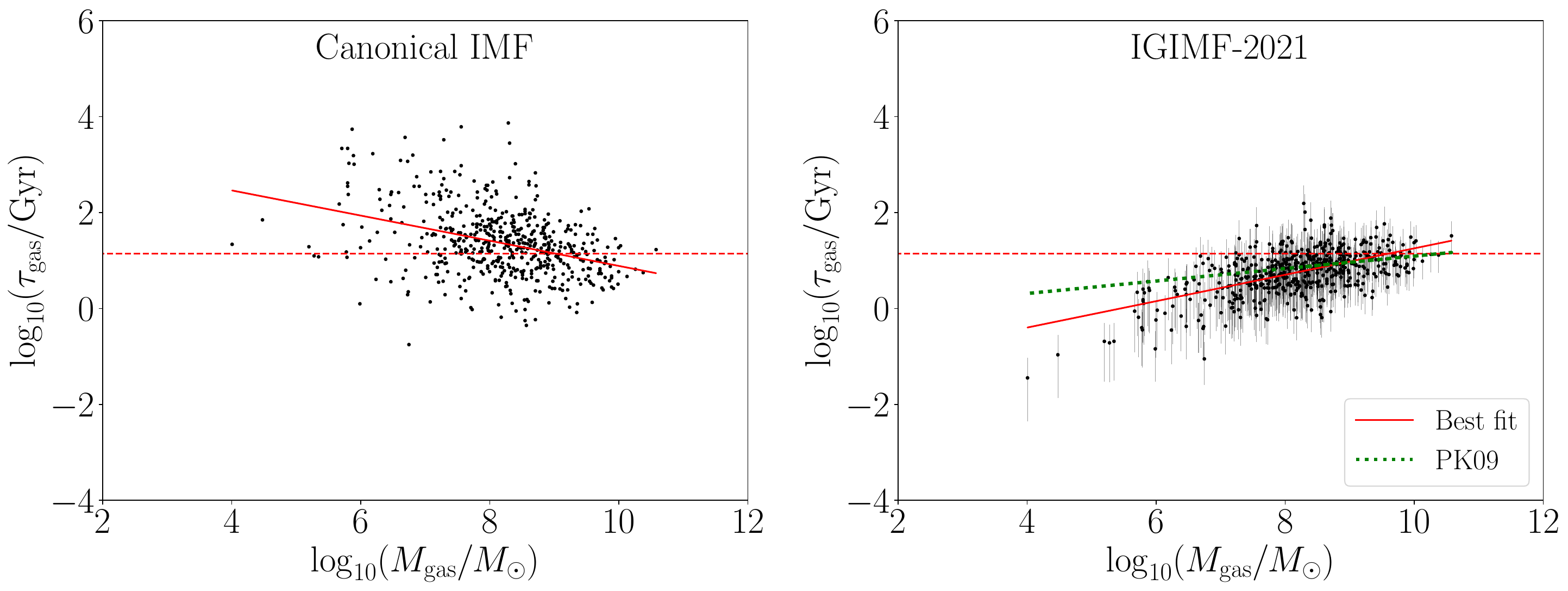}
    \caption{Gas depletion timescale, $\tau_{\mathrm{gas}}$, in dependence of the gas mass assuming the gwIMF in the invariant canonical IMF (left panel) and IGIMF-2021 (right panel) for galaxies located in the LV. The dashed red line shows the age of the Universe, $\tau_{\mathrm{gas}} = \tau_{\mathrm{h}} = 13.8 \, \rm{Gyr}$, assuming the standard $\Lambda$CDM cosmology \citep{Planck_2016_IllustrisTNG}. The error on the IGIMF-corrected $\tau_{\mathrm{gas}}$ values (right panel) is estimated by adopting the SFR values for $[Z] = -2.20$ (upper limit) and $[Z] = 0.45$ (lower limit; Section~\ref{subsec:M/L--L relation}). The solid red lines are fits of the data with equation~(\ref{eq:gasdepletion_timescale_Mgas_fit}). The green dotted line is the `minimum-1' model of \citet[their figure 7]{PflammAltenburg_2009b}. Note the smaller scatter of the data in the right panel (see also Fig.~\ref{figure_gasdepletion_timescale_Distribution}).}
    \label{figure_gasdepletion_timescale_IGIMFfit}
\end{figure*}

The present-day gas depletion timescale \citep[c.f. Section~5 of][]{PflammAltenburg_2009b} is defined by
\begin{eqnarray}
    \centering
    \tau_{\mathrm{gas}}~=~\frac{M_{\mathrm{gas}}}{\psi} \, ,
    \label{eq:gasdepletion_timescale}
\end{eqnarray}
such that the inverse is a measure of the SFE of a galaxy. The gas depletion timescales in dependence of the total neutral gas mass for the invariant canonical gwIMF and IGIMF-2021 for galaxies in the Local Cosmological Volume are presented in Fig.~\ref{figure_gasdepletion_timescale_IGIMFfit}. In the invariant canonical gwIMF context, the gas depletion timescale decreases with increasing gas mass, implying that low-mass galaxies have a lower SFE compared to more massive galaxies. The correlation is reversed assuming the gwIMF is given by the IGIMF-2021, in which low-mass galaxies have a higher SFE than massive galaxies. This is probably due to massive late-type galaxies having stronger feedback associated with the forming stellar population \citep[top-heavy gwIMF; see also][]{Gunawardhana_2011}. For the low-mass galaxies, the metallicity-dependent feedback is weaker in a low-metallicity environment which allows a higher SFE \citep{2011MNRAS.415.3439D,2011ApJ...737L..20D}.

Following \citet{PflammAltenburg_2009b}, the relation between the gas depletion timescale and total neutral gas is fitted with a function of the form
\begin{eqnarray}
    \centering
    \tau_{\mathrm{gas}}(M_{\mathrm{gas}})~=~A_{\mathrm{g}}~\bigg(\frac{M_{\rm{gas}}}{M_{\odot}} \bigg)^{B_{\mathrm{g}}} \, ,
    \label{eq:gasdepletion_timescale_Mgas_fit}
\end{eqnarray}
with $A_{\mathrm{g}} = 0.03$~Gyr and $B_{\mathrm{g}} = 0.28$ for the IGIMF-2021 case (solid red line in Fig.~\ref{figure_gasdepletion_timescale_IGIMFfit}) broadly matching the `minimum-1' model of \citet[their figure 7]{PflammAltenburg_2009b} with $A_{\mathrm{g}} = 0.62$~Gyr and $B_{\mathrm{g}} = 0.13$ (green dotted line in Fig.~\ref{figure_gasdepletion_timescale_IGIMFfit}) but significantly differs from their `standard' IGIMF model with $A_{\mathrm{g}} = 2.52$~Gyr and $B_{\mathrm{g}} = 0.01$ (see their equation~(7) and Fig.~6) implies only a weak correlation between $\tau_{\mathrm{gas}}$ and $M_{\rm{gas}}$.

\begin{figure}[!hbt]
    \includegraphics[width=\columnwidth]{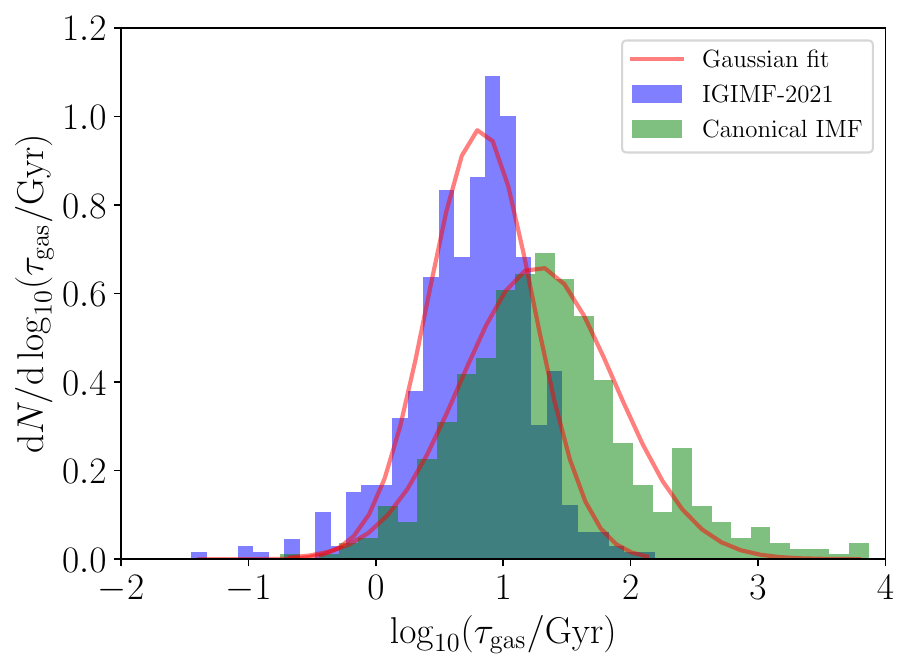}
    \caption{Distribution of the gas depletion timescale, $\tau_{\mathrm{gas}}$, assuming the gwIMF is given by the invariant canonical IMF (blue) and the IGIMF-2021 (green) in the $\log_{10}$-space. The $\tau_{\mathrm{gas}}$ are calculated by adopting the IGIMF-corrected SFR values for a Solar metallicity. The histograms are fitted with a Gaussian (solid red curves) with expected values of $\mu = \log_{10}(\tau_{\mathrm{gas}}/\rm{Gyr}) = 1.12$ (13.25~Gyr; canonical gwIMF) and $\mu = \log_{10}(\tau_{\mathrm{gas}}/\rm{Gyr}) = 0.67$ (4.72~Gyr; IGIMF-2021) and variances of $\sigma = 0.60$~Gyr (canonical gwIMF) and $\sigma = 0.41$~Gyr (IGIMF-2021).}
    \label{figure_gasdepletion_timescale_Distribution}
\end{figure}

Figure~\ref{figure_gasdepletion_timescale_Distribution} shows the distribution of the gas depletion timescales fitted with a log-normal function of the form \citep[following equation~(8) of][]{PflammAltenburg_2009b}
\begin{eqnarray}
    \centering
    \frac{{\rm d} N_{\mathrm{gal}}}{{\rm d} \log_{10}( \tau_{\mathrm{gas}})}~=~\frac{1}{\sqrt{2 \pi \sigma^{2}}} \exp \bigg( \frac{(\log_{10}(\tau_{\mathrm{gas}}) - \mu)^2}{2 \sigma^2} \bigg) \, ,
    \label{eq:gasdepletion_timescale_fit}
\end{eqnarray}
where $\mu$ and $\sigma$ are the mean and standard deviation, respectively, being $\mu = 1.12$ (13.25~Gyr) and $\sigma = 0.60$~Gyr in the case of the invariant canonical gwIMF and $\mu = 0.67$ (4.72~Gyr) and $\sigma = 0.41$~Gyr for the IGIMF-2021. The mean value but also the scatter of $\tau_{\mathrm{gas}}$ become smaller in the IGIMF framework, in agreement with fig.~8 of \citet{PflammAltenburg_2009b}.

\subsection{Characteristic stellar-mass buildup times} \label{subsec:Stellar-mass buildup times}
\begin{figure*}[!hbt]
    \includegraphics[width=\linewidth]{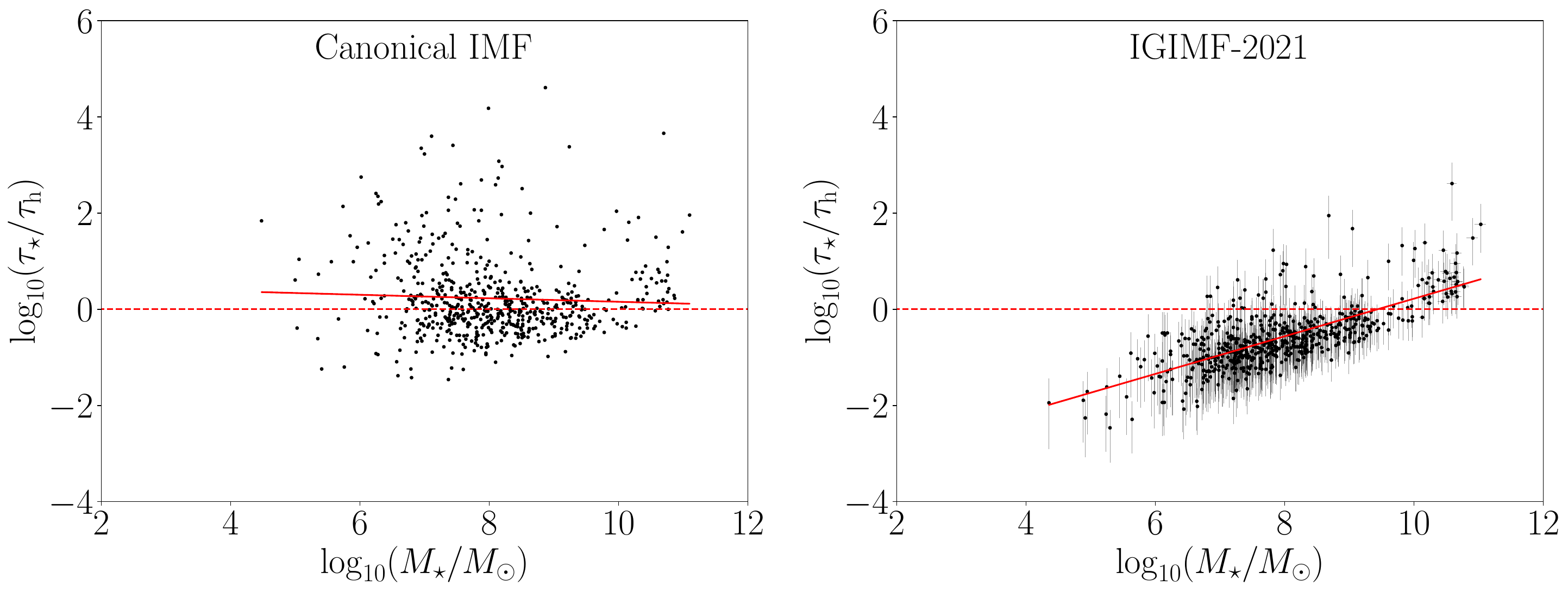}
    \caption{Characteristic stellar mass buildup timescale normalized by the age of the Universe of $\tau_{\mathrm{h}} = 13.8 \, \rm{Gyr}$ \citep{Planck_2016_IllustrisTNG} in dependence of the stellar mass in the invariant canonical IMF (left panel) and the IGIMF-2021 (right panel) context for galaxies located in the LV. The dashed red line refers to a characteristic stellar mass buildup timescale equal to $\tau_{\mathrm{h}} = 13.8 \, \rm{Gyr}$. In the right panel, the uncertainty of the IGIMF effect is plotted while the measurement errors are not shown. The upper and lower limits of the IGIMF-corrected $\tau_\star$ values (right panel) are estimated by adopting $\tau_{\mathrm{up},\star} = M_{\mathrm{upper},\star}/\psi([Z] = -2.20)$ and $\tau_{\mathrm{low},\star} = M_{\mathrm{low},\star}/\psi([Z] = 0.45)$, respectively. $\psi([Z] = -2.20)$ and $\psi([Z] = 0.45)$ refer to the IGIMF-corrected SFR values for $[Z] = -2.20$ and $[Z] = 0.45$ (Section~\ref{subsec:M/L--L relation}). $M_{\mathrm{up},\star}$ and $M_{\mathrm{low},\star}$ refer to the upper and lower IGIMF-corrected stellar masses by applying the maximum and minimum $M_\star/L_{\mathrm{Ks}}$ values (Section~\ref{subsec:M/L--L relation}). The solid red lines are fits of the data with equation~(\ref{eq:stellarmassbuildupfit}).
    }
    \label{figure_stellarmass_buildup_timescale}
\end{figure*}

The characteristic stellar mass buildup timescale \citep[Section~6 of][]{PflammAltenburg_2009b} is defined as
\begin{eqnarray}\label{eq stellar mass buildup timescale}
    \centering
    \tau_\star~=~\frac{M_\star}{\psi}.
    \label{eq:tau_star}
\end{eqnarray}
In the invariant canonical gwIMF context, the characteristic stellar mass buildup timescale decreases with increasing stellar mass of the galaxies as shown in the left panel of Fig.~\ref{figure_stellarmass_buildup_timescale}. In particular, low-mass galaxies with present-day stellar mass $M_\star \la 10^{8} \,M_{\odot}$ have $\tau_\star$ values much larger than the age of the Universe $\tau_{\mathrm{h}} = 13.8$~Gyr, implying that they have a lower than average SFR at the present day, while more massive galaxies have $\tau_\star \approx \tau_{\mathrm{h}}$, that is, $\psi \approx M_\star/\tau_{\mathrm{h}}$. 

In contrast, the characteristic stellar mass buildup timescale increases with stellar mass in the IGIMF-2021 context.
Fitting the relation between the stellar-mass buildup timescale normalized by $\tau_{\mathrm{h}}$ and the stellar mass with
\begin{eqnarray}
    \centering
    \frac{\tau_\star(M_\star)}{\tau_{\mathrm{h}}}=~A_\star~\bigg(\frac{M_\star}{M_{\odot}} \bigg)^{B_\star} \, ,
    \label{eq:stellarmassbuildupfit}
\end{eqnarray}
yields $A_\star = 3.86\times 10^{-4}$ and $B_\star = 0.37$ for the IGIMF-2021 case (solid red line in Fig.~\ref{figure_stellarmass_buildup_timescale}). 
Most of the galaxies with $M_\star \la 10^{10}\,M_{\odot}$ have $\tau_\star < \tau_{\mathrm{h}}$ but more massive galaxies have $\tau_\star > \tau_{\mathrm{h}}$.
Intriguingly, we see again that the scatter of the galaxies around this average relation is much tighter than the case when assuming the invariant canonical gwIMF, revealing a regulated formation history of galaxies located in the LV.
This result is in agreement with the `minimum-1' IGIMF formulation in \citet{PflammAltenburg_2009b} and is discussed further in Section~\ref{subsec:StellarBuildup_Discussion}.

\section{Discussion}
\label{sec: Discussion}
Following Section~4.1 of \citet{Jerabkova_2018}, we present in Section~\ref{subsec:IGIMF theory} a grid of gwIMFs in dependence of the averaged metallicity and global SFR constructed from the IGIMF-2021 formulation \citep{Yan_2021}. The gwIMF systematically varies with metallicity and SFR in such a way that it becomes top-heavy for galaxies with $\psi > \mathrm{few} \,~\rm{\, M_{\odot}\,yr^{-1}}$, while being top-light when $\psi < 1\,~\rm{\, M_{\odot}\,yr^{-1}}$ compared to the invariant canonical gwIMF. The gwIMF becomes bottom-heavy and bottom-light for [Z] > 0 and [Z] < 0, respectively, independent of the SFR. The gwIMF is close to a canonical IMF when $\psi = 1 \, \rm{M_{\odot}\,yr^{-1}}$ and at solar metallicity (Section~\ref{subsec:IGIMF theory}). The gwIMF variation for the most metal-rich galaxies will be discussed in \citet{Yan24} and Gjergo et al. (in preparation).

\subsection{Stellar-mass--SFR relation of star-forming galaxies for a varying IMF} \label{sec:MainSequence_Discussion}
As the Kennicutt-law is invalid for a varying gwIMF, the stellar-mass--SFR relation and therefore also the main sequence of star-forming galaxies needs to be revised within the current parameterization of the IGIMF framework. We present the SFR--H$\alpha$ luminosity relation in the IGIMF-2021 formulation for different metallicities using the P\'{E}GASE stellar population synthesis code (Section~\ref{subsec: SFR-Halpha relation}) and provide fitting functions in order to transform observed H$\alpha$ luminosities or Kennicutt SFRs to IGIMF SFRs  (Appendix~\ref{appendix_fittingparameters}). 

For example, less luminous H$\alpha$ galaxies with $L_{\mathrm{H\alpha}} = 5.5 \times 10^{36}\,\rm{erg\,s^{-1}}$ \citep[e.g. the dwarf galaxy Leo P;][]{Jerabkova_2018} have SFRs about 15 (for [Z] = -2.20; Section~\ref{subsec:M/L--L relation}) to 221 (for [Z] = 0.45) times higher than based on the Kennicutt law, while more H$\alpha$-luminous galaxies with $L_{\mathrm{H\alpha}} = 10^{44}\,\rm{erg\,s^{-1}}$ have about 4.4 (for [Z] = -2.20) and 13 (for [Z] = 0.45) times lower SFRs. As a consequence, the IGIMF-corrected present-day main sequence of galaxies has a shallower slope (Fig.~\ref{figure_main_sequence_IGIMF_observations}).  

\citet{Kroupa_Jerabkova_2021} studied the redshift evolution of the main sequence by only correcting the SFRs and also found a shallower slope and a weaker redshift dependence. Thus, the IGIMF theory has also cosmological implications. Interestingly, several studies found inconsistencies between the observed and stellar mass density and SFR density over cosmic time \citep[e.g.][]{Grazian_2015,Hopkins_2006,Hopkins_2008,Wilkins_2008,Madau_Dickinson_2014,Leja_2015,Tomczak_2016,Yu_2016}. A top-heavy or bottom-light gwIMF could alleviate this discrepancy \citep{Yu_2016}. However, such an analysis is beyond the scope of our article and depends on the physical interpretation of the \citet{Madau_Dickinson_2014} data of SFR density versus redshift $z$ \citep{Haslbauer_2023a}. Given the near-constant SFRs for the galaxies observed in the LV these data rather imply a significant matter over-density at $z\approx 1.8$.

\subsection{Gas depletion timescales} \label{subsec:GasDepleation_Discussion}
\citet{PflammAltenburg_2009b} applied a revised SFR--H$\alpha$ luminosity relation based on 200 local galaxies for IGIMF versions which assume a canonical IMF per embedded star cluster and an ECMF with a single power-law index of $\beta = 2.35$ (their `standard' IGIMF model) and a two-part power law with $\beta = 1.00$ for $5 < m/M_{\odot} < 50$ and $\beta = 2.00$ for $m/M_{\odot} > 50$ (`minimum-1' IGIMF model). Their standard model shows a stronger IGIMF effect compared to the `minimum-1' model by yielding higher SFRs (see their Fig.~1).

In the invariant canonical gwIMF case, the gas depletion timescale shows a decreasing trend with increasing gas mass implying that less massive galaxies have a lower SFE \citep{PflammAltenburg_2009b}. That is, galaxies with $M_{\mathrm{gas}} \la 10^{9}\,M_{\odot}$ have $\tau_{\mathrm{gas}}$ up to $10^{4}~\rm{Gyr}$ while more massive galaxies have $\tau_{\mathrm{gas}} < 10^{2}$~Gyr. In the IGIMF-2021 case, the situation differs as $\tau_{\mathrm{gas}}$ slightly increases with gas mass such that low-mass galaxies have a higher SFE (i.e. shorter $\tau_{\mathrm{gas}}$ values). Almost all galaxies have $\tau_{\mathrm{gas}} \la 10^{2}$~Gyr with a mean of a few Gyr (Fig.~\ref{figure_gasdepletion_timescale_Distribution}). The scatter of $\tau_{\mathrm{gas}}$ is smaller assuming the IGIMF-2021 than for the invariant canonical gwIMF. 

These findings are consistent with the `minimum-1' IGIMF formalism of \citet{PflammAltenburg_2009b} while their standard IGIMF yields an almost constant gas depletion timescale between $2.5$ and $3$~Gyr over a stellar mass range of $\approx10^{6}~{\rm to}~10^{10}~M_{\odot}$. 

The result obtained here of a decreasing SFE with increasing gas (and thus galaxy) mass may suggest that dwarf late-type galaxies form more stars per unit gas mass per unit of time because their star-formation process occurs mostly in low-mass embedded clusters that do not have cloud-destroying ionizing stars.

\subsection{Stellar-mass buildup timescales} \label{subsec:StellarBuildup_Discussion}
In the invariant canonical gwIMF case, the characteristic stellar mass buildup timescale decreases with increasing stellar mass for galaxies located in the LV. The situation changes in the IGIMF-2021 context as the characteristic stellar mass buildup timescale increases with stellar mass. Thus, our findings, with a more self-consistent galaxy mass estimation in the IGIMF framework, qualitatively confirm the results of \citet{PflammAltenburg_2009b} for a larger sample size of local galaxies, suggesting that massive star-forming galaxies have a relatively constant or decreasing SFR, while dwarf galaxies have either started their star-formation activity recently or have an increasing SFR, as is indeed observed to be the case for the galaxies in the Local Group \citep{Kroupa_2020, Ren_2024}. Compared to the results assuming an invariant canonical IMF, the dependency of the SFH on the stellar mass of a galaxy calculated using the IGIMF-2021 theory agrees better with simulation and population synthesis studies (e.g. \citealt{2014MNRAS.444.2637M,2015MNRAS.450.2749G,2023MNRAS.526.1022L}).

For completeness we note that the characteristic stellar mass buildup timescale, calculated according to equation~(\ref{eq stellar mass buildup timescale}), may not be the real stellar mass buildup time of a galaxy because i) the present-day stellar mass, $M_\star$, may not be the total stellar mass ever formed in a galaxy and ii) the present-day SFR estimation based on the H$\alpha$ luminosity may not be the same as the mean SFR of a galaxy.
Not all galaxies have a relatively stable SFR. For example, \citet{Kroupa_2020} reported that most of the galaxies located in the LV have nearly constant SFRs over a star-formation timescale of $12$~Gyr, while \citet{Speagle_2014} showed that massive star-forming galaxies with $M_\star\ga 10^{10}\,M_{\odot}$ strictly evolving along the main sequence of galaxies can be described by a delayed-tau SFH model with maximum SFRs when the galaxy was $3.5-4.5$~Gyr old. Re-addressing the analysis for more complex SFHs would be required in upcoming studies.

It is worth addressing the impact that galaxy mergers may have on the analysis of SFHs. In this context, however, their influence is generally deemed negligible for several reasons. First, we note that the vast majority of galaxies are thin disk galaxies  \citep{Binggeli_1988, DelgadoSerrano_2010, Haslbauer_2022_morphology}, whose evolution is best characterized by in-situ star formation \citep[e.g.,][]{Zonoozi_2019}. This is also evident in the Milky Way \citep[e.g.,][]{Belokurov24, ChenGnedin24}. Therefore, our model is well suited to describe spiral galaxies. But even within a hierarchical mass assembly framework, our model remains applicable. In $\Lambda$CDM cosmological simulations \citep[e.g.][]{2013ApJ...766...38L} and observations \citep{2022A&A...664A.115Z} galaxies similar to or less massive than the Milky Way have a small fraction ($\lesssim 20\%$) of merged stars, and our galaxy sample does not exceed this mass range. 
Ultimately, a comprehensive SFH analysis must include dynamical evolution and hence require the self-consistency of cosmological simulations, paired with stellar population synthesis studies. However, we advise caution in the excessive reliance on hierarchical mass assembly models, which lead to inconsistencies with observations. For example, \citealt{Eappen_2022} revealed that the stellar age distributions in hierarchically-formed elliptical galaxies span too wide an age range, in contrast with observations. Additionally, \citealt{Haslbauer_2022_morphology} noted that $\Lambda$CDM-based simulations overproduce elliptical galaxies, contradicting the observation that the universe predominantly contains spirals.

Therefore, it would be valuable to implement the IGIMF theory in cosmological simulations. Attempts towards incorporating a varying IMF in Newtonian simulations have been made for example by \citet[][]{Ploeckinger_2014} and \citet{Ploeckinger_2015}, who encoded the IGIMF theory in the FLASH code. This would allow us to revisit downsizing in self-consistent cosmological simulations of structure formation and to address if a top-heavy gwIMF could indeed explain the JWST observations of very luminous galaxy candidates at (photometric) redshifts $z> 10$ within the standard $\Lambda$CDM paradigm \citep{Haslbauer_2022b}. 
Implementing the IGIMF theory in the Phantom of Ramses \citep[PoR MOND patch developed by][and the corresponding user guide by \citealt{Nagesh_2021}]{Lueghausen_2015} will enable galaxy evolution in Milgromian dynamics \citep[MOND,][]{Milgrom_1983a} where mergers are rare \citep{Kroupa_2023}. 

\section{Conclusion} \label{sec: Conclusion}

A systematically varying gwIMF has fundamental implications on the global properties of the galaxy as the stellar population is constructed from the IMF. In this contribution, we investigate the effect of the IGIMF theory, which is a mathematical framework to compute the gwIMF by integrating all IMFs of star-forming regions (embedded clusters) in a galaxy (Section~\ref{subsec:IGIMF theory}), on the estimation of galaxy mass and SFR for a sample of local star-forming galaxies (Section~\ref{subsec: Data}).

To convert infrared luminosities to stellar masses, we developed the photGalIMF code to calculate the evolution of $M_\star/L$ (including remnants) in the Ks-band (centred at $\approx 2.2~\mu\rm m$) in a galaxy chemical evolution model with an environment-dependent gwIMF according to the IGIMF theory (Section~\ref{subsubsec:stellar evolution}). The photGalIMF code provides at its current stage only the Ks-band, the IRAC [3.6]-band, and the V-band and will include further bands such as the \textit{Gaia} and JWST photometries in the future.

The SFR--H$\alpha$-luminosity relation for different metallicities is calculated with the P\'{E}GASE stellar population synthesis code. We provided new correcting functions for converting the measured H$\alpha$ luminosity or SFRs derived from the Kennicutt-law to SFRs interpreted within the IGIMF-2021 context. The estimated SFR under these two assumptions can be different by a factor of 100 (Section~\ref{subsec: SFR-Halpha relation}). In the future, hydrogen line luminosities should be added to the photGalIMF code to allow a self-consistent calculation of the SFR--H$\alpha$ luminosity relation in galaxy evolution simulations for different SFHs and chemical enrichment models. 

Assessing the gwIMF in the framework of the IGIMF-2021 formulation, we find that
\begin{itemize}
\item The $M_{\star}/L$ ratio of a galaxy depends significantly on which description of the IMF (invariant canonical or the IGIMF) is applied (Section~\ref{subsec:Stellar mass-to-light ratio} and \ref{subsec:M/L--L relation}). 

\item The SFR--stellar-mass relation of star-forming galaxies becomes flatter and has a smaller scatter compared to the results obtained with the canonical invariant IMF, extending the main sequence to dwarf galaxies (Section~\ref{subsec:main sequence of star-forming galaxies}). 

\item The gas depletion timescale of local star-forming galaxies increases with their total gas mass in the IGIMF context, suggesting a lower SFE (i.e. longer gas-consumption timescales) in massive galaxies (Section~\ref{subsec:Gas depletion timescale}). 

\item The characteristic stellar mass buildup timescale of local star-forming galaxies increases with their stellar masses, indicating that massive star-forming galaxies have on average a relatively constant or slightly decreasing SFR while dwarf galaxies either have an increasing SFR or that they started their star formation more recently.
\end{itemize}

Interpreting photometric observations and subsequently testing cosmological and galaxy evolution models requires assumptions about the stellar IMF. The here-presented photGalIMF code enables a more accurate comparison of photometric observations with galaxy evolution and cosmological models.

\begin{acknowledgements}
    The authors thank Alice Concas, Zhi-Yu Zhang, and Federico Lelli for helpful discussions. M.H. acknowledges support from the ESO Early-Career Visitor Programme and the SPODYR group at Bonn University for a studentship. Z.Y. acknowledges the support from the Jiangsu Funding Program for Excellent Postdoctoral Talent under grant number 2022ZB54, the National Natural Science Foundation of China under grant numbers 12203021, 12041305, and 12173016, and the Fundamental Research Funds for the Central Universities under grant number 0201/14380049. The authors thank an anonymous referee for very helpful comments. The data underlying this article are available in the article. The Python~3 code GalIMF\footref{link:Github_galIMF} version 1.1 developed by \citet{Yan_2017,Yan_2019a} with the package photGalIMF\footref{link:Github_photgalIMF} version 1.0 (Section~\ref{subsubsec:stellar evolution}) is publicly available on GitHub. The observational data are taken from the updated version of the Catalogue of Neighbouring Galaxies \citep{Karachentsev_2004, Karachentsev_2013} and have been IGIMF-corrected as described in Section~\ref{subsec:M/L--L relation}.
\end{acknowledgements}

\bibliographystyle{aa} 
\bibliography{references} 

\begin{thebibliography}{169}
\expandafter\ifx\csname natexlab\endcsname\relax\def\natexlab#1{#1}\fi

\bibitem[{{Asplund} {et~al.}(2009){Asplund}, {Grevesse}, {Sauval}, \&
  {Scott}}]{Asplund_2009}
{Asplund}, M., {Grevesse}, N., {Sauval}, A.~J., \& {Scott}, P. 2009, \araa, 47,
  481

\bibitem[{{Balakrishna Subramani} {et~al.}(2019){Balakrishna Subramani},
  {Kroupa}, {Shenavar}, \& {Muralidhara}}]{BalakrishnaSubramani_2019}
{Balakrishna Subramani}, V., {Kroupa}, P., {Shenavar}, H., \& {Muralidhara}, V.
  2019, \mnras, 488, 3876

\bibitem[{{Bell} \& {de Jong}(2001)}]{2001ApJ...550..212B}
{Bell}, E.~F. \& {de Jong}, R.~S. 2001, \apj, 550, 212

\bibitem[{{Belokurov} \& {Kravtsov}(2024)}]{Belokurov24}
{Belokurov}, V. \& {Kravtsov}, A. 2024, \mnras, 528, 3198

\bibitem[{{Bessell}(1990)}]{Bessell_1990}
{Bessell}, M.~S. 1990, \pasp, 102, 1181

\bibitem[{{Binggeli} {et~al.}(1988){Binggeli}, {Sandage}, \&
  {Tammann}}]{Binggeli_1988}
{Binggeli}, B., {Sandage}, A., \& {Tammann}, G.~A. 1988, \araa, 26, 509

\bibitem[{{Bressan} {et~al.}(2012){Bressan}, {Marigo}, {Girardi}, {Salasnich},
  {Dal Cero}, {Rubele}, \& {Nanni}}]{Bressan_2012}
{Bressan}, A., {Marigo}, P., {Girardi}, L., {et~al.} 2012, \mnras, 427, 127

\bibitem[{{Brinchmann} {et~al.}(2004){Brinchmann}, {Charlot}, {White},
  {Tremonti}, {Kauffmann}, {Heckman}, \& {Brinkmann}}]{Brinchmann_2004}
{Brinchmann}, J., {Charlot}, S., {White}, S.~D.~M., {et~al.} 2004, \mnras, 351,
  1151

\bibitem[{{Brown} {et~al.}(2017){Brown}, {Moustakas}, {Kennicutt}, {Bonne},
  {Intema}, {de Gasperin}, {Boquien}, {Jarrett}, {Cluver}, {Smith}, {da Cunha},
  {Imanishi}, {Armus}, {Brandl}, \& {Peek}}]{Brown_2017}
{Brown}, M. J.~I., {Moustakas}, J., {Kennicutt}, R.~C., {et~al.} 2017, \apj,
  847, 136

\bibitem[{{Buat} {et~al.}(1989){Buat}, {Deharveng}, \& {Donas}}]{Buat_1989}
{Buat}, V., {Deharveng}, J.~M., \& {Donas}, J. 1989, \aap, 223, 42

\bibitem[{{Busch} {et~al.}(2015){Busch}, {Smaji{\'c}}, {Scharw{\"a}chter},
  {Eckart}, {Valencia-S.}, {Moser}, {Husemann}, {Krips}, \&
  {Zuther}}]{2015A&A...575A.128B}
{Busch}, G., {Smaji{\'c}}, S., {Scharw{\"a}chter}, J., {et~al.} 2015, \aap,
  575, A128

\bibitem[{{Calzetti}(2013)}]{Calzetti_2013}
{Calzetti}, D. 2013, in Secular Evolution of Galaxies, ed.
  J.~{Falc{\'o}n-Barroso} \& J.~H. {Knapen}, 419

\bibitem[{{Cameron} {et~al.}(2023){Cameron}, {Katz}, {Witten}, {Saxena},
  {Laporte}, \& {Bunker}}]{2023arXiv231102051C}
{Cameron}, A.~J., {Katz}, H., {Witten}, C., {et~al.} 2023, arXiv e-prints,
  arXiv:2311.02051

\bibitem[{{Chen} \& {Gnedin}(2024)}]{ChenGnedin24}
{Chen}, Y. \& {Gnedin}, O.~Y. 2024, The Open Journal of Astrophysics, 7, 23

\bibitem[{{Cohen} {et~al.}(2003){Cohen}, {Wheaton}, \& {Megeath}}]{Cohen_2003}
{Cohen}, M., {Wheaton}, W.~A., \& {Megeath}, S.~T. 2003, \aj, 126, 1090

\bibitem[{{Concas} {et~al.}(2022){Concas}, {Maiolino}, {Curti},
  {Hayden-Pawson}, {Cirasuolo}, {Jones}, {Mercurio}, {Belfiore}, {Cresci},
  {Cullen}, {Mannucci}, {Marconi}, {Cappellari}, {Cicone}, {Peng}, \&
  {Troncoso}}]{Concas_2022}
{Concas}, A., {Maiolino}, R., {Curti}, M., {et~al.} 2022, \mnras, 513, 2535

\bibitem[{{Condon}(1992)}]{Condon_1992}
{Condon}, J.~J. 1992, \araa, 30, 575

\bibitem[{{Dabringhausen} {et~al.}(2009){Dabringhausen}, {Kroupa}, \&
  {Baumgardt}}]{Dabringhausen_2009}
{Dabringhausen}, J., {Kroupa}, P., \& {Baumgardt}, H. 2009, \mnras, 394, 1529

\bibitem[{{Dabringhausen} {et~al.}(2012){Dabringhausen}, {Kroupa},
  {Pflamm-Altenburg}, \& {Mieske}}]{Dabringhausen_2012}
{Dabringhausen}, J., {Kroupa}, P., {Pflamm-Altenburg}, J., \& {Mieske}, S.
  2012, \apj, 747, 72

\bibitem[{{David} {et~al.}(1992){David}, {Jones}, \& {Forman}}]{David_1992}
{David}, L.~P., {Jones}, C., \& {Forman}, W. 1992, \apj, 388, 82

\bibitem[{{Delgado-Serrano} {et~al.}(2010){Delgado-Serrano}, {Hammer}, {Yang},
  {Puech}, {Flores}, \& {Rodrigues}}]{DelgadoSerrano_2010}
{Delgado-Serrano}, R., {Hammer}, F., {Yang}, Y.~B., {et~al.} 2010, \aap, 509,
  A78

\bibitem[{{Dib}(2011)}]{2011ApJ...737L..20D}
{Dib}, S. 2011, \apjl, 737, L20

\bibitem[{{Dib}(2014)}]{2014MNRAS.444.1957D}
{Dib}, S. 2014, \mnras, 444, 1957

\bibitem[{{Dib}(2022)}]{2022A&A...666A.113D}
{Dib}, S. 2022, \aap, 666, A113

\bibitem[{{Dib}(2023)}]{2023ApJ...959...88D}
{Dib}, S. 2023, \apj, 959, 88

\bibitem[{{Dib} \& {Basu}(2018)}]{2018A&A...614A..43D}
{Dib}, S. \& {Basu}, S. 2018, \aap, 614, A43

\bibitem[{{Dib} {et~al.}(2011){Dib}, {Piau}, {Mohanty}, \&
  {Braine}}]{2011MNRAS.415.3439D}
{Dib}, S., {Piau}, L., {Mohanty}, S., \& {Braine}, J. 2011, \mnras, 415, 3439

\bibitem[{{Dib} {et~al.}(2017){Dib}, {Schmeja}, \& {Hony}}]{Dib_2017}
{Dib}, S., {Schmeja}, S., \& {Hony}, S. 2017, \mnras, 464, 1738

\bibitem[{{Dinnbier} {et~al.}(2022){Dinnbier}, {Kroupa}, \&
  {Anderson}}]{Dinnbier_2022}
{Dinnbier}, F., {Kroupa}, P., \& {Anderson}, R.~I. 2022, \aap, 660, A61

\bibitem[{{Donas} \& {Deharveng}(1984)}]{Donas_1984}
{Donas}, J. \& {Deharveng}, J.~M. 1984, \aap, 140, 325

\bibitem[{{Eappen} {et~al.}(2022){Eappen}, {Kroupa}, {Wittenburg}, {Haslbauer},
  \& {Famaey}}]{Eappen_2022}
{Eappen}, R., {Kroupa}, P., {Wittenburg}, N., {Haslbauer}, M., \& {Famaey}, B.
  2022, \mnras, 516, 1081

\bibitem[{{Fioc} {et~al.}(2011){Fioc}, {Le Borgne}, \&
  {Rocca-Volmerange}}]{Fioc_2011}
{Fioc}, M., {Le Borgne}, D., \& {Rocca-Volmerange}, B. 2011, {P{\'E}GASE:
  Metallicity-consistent Spectral Evolution Model of Galaxies}, Astrophysics
  Source Code Library, record ascl:1108.007

\bibitem[{{Fioc} \& {Rocca-Volmerange}(1999)}]{Fioc_PEGASE.2_1999}
{Fioc}, M. \& {Rocca-Volmerange}, B. 1999, arXiv e-prints, astro

\bibitem[{{Fontanot} {et~al.}(2009){Fontanot}, {De Lucia}, {Monaco},
  {Somerville}, \& {Santini}}]{Fontanot_2009}
{Fontanot}, F., {De Lucia}, G., {Monaco}, P., {Somerville}, R.~S., \&
  {Santini}, P. 2009, \mnras, 397, 1776

\bibitem[{{Fontanot} {et~al.}(2023){Fontanot}, {La Barbera}, {De Lucia},
  {Cecchi}, {Xie}, {Hirschmann}, {Bruzual}, {Charlot}, \&
  {Vazdekis}}]{2023arXiv231112932F}
{Fontanot}, F., {La Barbera}, F., {De Lucia}, G., {et~al.} 2023, arXiv
  e-prints, arXiv:2311.12932

\bibitem[{{Fontanot} {et~al.}(2018){Fontanot}, {La Barbera}, {De Lucia},
  {Pasquali}, \& {Vazdekis}}]{Fontanot_2018}
{Fontanot}, F., {La Barbera}, F., {De Lucia}, G., {Pasquali}, A., \&
  {Vazdekis}, A. 2018, \mnras, 479, 5678

\bibitem[{{F{\"o}rster Schreiber} \& {Wuyts}(2020)}]{2020ARA&A..58..661F}
{F{\"o}rster Schreiber}, N.~M. \& {Wuyts}, S. 2020, \araa, 58, 661

\bibitem[{{Fukugita} \& {Peebles}(2004)}]{2004ApJ...616..643F}
{Fukugita}, M. \& {Peebles}, P.~J.~E. 2004, \apj, 616, 643

\bibitem[{{Gallazzi} {et~al.}(2005){Gallazzi}, {Charlot}, {Brinchmann},
  {White}, \& {Tremonti}}]{Gallazzi_2005}
{Gallazzi}, A., {Charlot}, S., {Brinchmann}, J., {White}, S. D.~M., \&
  {Tremonti}, C.~A. 2005, \mnras, 362, 41

\bibitem[{{Geha} {et~al.}(2013){Geha}, {Brown}, {Tumlinson}, {Kalirai},
  {Simon}, {Kirby}, {VandenBerg}, {Mu{\~n}oz}, {Avila}, {Guhathakurta}, \&
  {Ferguson}}]{Geha_2013}
{Geha}, M., {Brown}, T.~M., {Tumlinson}, J., {et~al.} 2013, \apj, 771, 29

\bibitem[{{Gennaro} {et~al.}(2018){Gennaro}, {Tchernyshyov}, {Brown}, {Geha},
  {Avila}, {Guhathakurta}, {Kalirai}, {Kirby}, {Renzini}, {Simon}, {Tumlinson},
  \& {Vargas}}]{Gennaro_2018}
{Gennaro}, M., {Tchernyshyov}, K., {Brown}, T.~M., {et~al.} 2018, \apj, 855, 20

\bibitem[{{Girardi} \& {Bertelli}(1998)}]{Girardi_1998}
{Girardi}, L. \& {Bertelli}, G. 1998, \mnras, 300, 533

\bibitem[{{Girardi} {et~al.}(2013){Girardi}, {Marigo}, {Bressan}, \&
  {Rosenfield}}]{Girardi_2013}
{Girardi}, L., {Marigo}, P., {Bressan}, A., \& {Rosenfield}, P. 2013, \apj,
  777, 142

\bibitem[{{Grazian} {et~al.}(2015){Grazian}, {Fontana}, {Santini}, {Dunlop},
  {Ferguson}, {Castellano}, {Amorin}, {Ashby}, {Barro}, {Behroozi}, {Boutsia},
  {Caputi}, {Chary}, {Dekel}, {Dickinson}, {Faber}, {Fazio}, {Finkelstein},
  {Galametz}, {Giallongo}, {Giavalisco}, {Grogin}, {Guo}, {Kocevski},
  {Koekemoer}, {Koo}, {Lee}, {Lu}, {Merlin}, {Mobasher}, {Nonino}, {Papovich},
  {Paris}, {Pentericci}, {Reddy}, {Renzini}, {Salmon}, {Salvato}, {Sommariva},
  {Song}, \& {Vanzella}}]{Grazian_2015}
{Grazian}, A., {Fontana}, A., {Santini}, P., {et~al.} 2015, \aap, 575, A96

\bibitem[{{Groenewegen}(2006)}]{Groenewegen_2006}
{Groenewegen}, M.~A.~T. 2006, \aap, 448, 181

\bibitem[{{Guglielmo} {et~al.}(2015){Guglielmo}, {Poggianti}, {Moretti},
  {Fritz}, {Calvi}, {Vulcani}, {Fasano}, \&
  {Paccagnella}}]{2015MNRAS.450.2749G}
{Guglielmo}, V., {Poggianti}, B.~M., {Moretti}, A., {et~al.} 2015, \mnras, 450,
  2749

\bibitem[{{Gunawardhana} {et~al.}(2011){Gunawardhana}, {Hopkins}, {Sharp},
  {Brough}, {Taylor}, {Bland-Hawthorn}, {Maraston}, {Tuffs}, {Popescu},
  {Wijesinghe}, {Jones}, {Croom}, {Sadler}, {Wilkins}, {Driver}, {Liske},
  {Norberg}, {Baldry}, {Bamford}, {Loveday}, {Peacock}, {Robotham}, {Zucker},
  {Parker}, {Conselice}, {Cameron}, {Frenk}, {Hill}, {Kelvin}, {Kuijken},
  {Madore}, {Nichol}, {Parkinson}, {Pimbblet}, {Prescott}, {Sutherland},
  {Thomas}, \& {van Kampen}}]{Gunawardhana_2011}
{Gunawardhana}, M.~L.~P., {Hopkins}, A.~M., {Sharp}, R.~G., {et~al.} 2011,
  \mnras, 415, 1647

\bibitem[{{Haslbauer} {et~al.}(2022{\natexlab{a}}){Haslbauer}, {Banik},
  {Kroupa}, {Wittenburg}, \& {Javanmardi}}]{Haslbauer_2022_morphology}
{Haslbauer}, M., {Banik}, I., {Kroupa}, P., {Wittenburg}, N., \& {Javanmardi},
  B. 2022{\natexlab{a}}, \apj, 925, 183

\bibitem[{{Haslbauer} {et~al.}(2023){Haslbauer}, {Kroupa}, \&
  {Jerabkova}}]{Haslbauer_2023a}
{Haslbauer}, M., {Kroupa}, P., \& {Jerabkova}, T. 2023, \mnras, 524, 3252

\bibitem[{{Haslbauer} {et~al.}(2022{\natexlab{b}}){Haslbauer}, {Kroupa},
  {Zonoozi}, \& {Haghi}}]{Haslbauer_2022b}
{Haslbauer}, M., {Kroupa}, P., {Zonoozi}, A.~H., \& {Haghi}, H.
  2022{\natexlab{b}}, \apjl, 939, L31

\bibitem[{{Hopkins}(2018)}]{Hopkins_2018}
{Hopkins}, A.~M. 2018, \pasa, 35, e039

\bibitem[{{Hopkins} \& {Beacom}(2006)}]{Hopkins_2006}
{Hopkins}, A.~M. \& {Beacom}, J.~F. 2006, \apj, 651, 142

\bibitem[{{Hopkins} \& {Beacom}(2008)}]{Hopkins_2008}
{Hopkins}, A.~M. \& {Beacom}, J.~F. 2008, \apj, 682, 1486

\bibitem[{{Hopkins} {et~al.}(2018){Hopkins}, {Wetzel}, {Kere{\v{s}}},
  {Faucher-Gigu{\`e}re}, {Quataert}, {Boylan-Kolchin}, {Murray}, {Hayward},
  {Garrison-Kimmel}, {Hummels}, {Feldmann}, {Torrey}, {Ma},
  {Angl{\'e}s-Alc{\'a}zar}, {Su}, {Orr}, {Schmitz}, {Escala}, {Sanderson},
  {Grudi{\'c}}, {Hafen}, {Kim}, {Fitts}, {Bullock}, {Wheeler}, {Chan},
  {Elbert}, \& {Narayanan}}]{2018MNRAS.480..800H}
{Hopkins}, P.~F., {Wetzel}, A., {Kere{\v{s}}}, D., {et~al.} 2018, \mnras, 480,
  800

\bibitem[{{Jarrett} {et~al.}(2000){Jarrett}, {Chester}, {Cutri}, {Schneider},
  {Skrutskie}, \& {Huchra}}]{Jarrett_2000}
{Jarrett}, T.~H., {Chester}, T., {Cutri}, R., {et~al.} 2000, \aj, 119, 2498

\bibitem[{{Jarrett} {et~al.}(2003){Jarrett}, {Chester}, {Cutri}, {Schneider},
  \& {Huchra}}]{Jarrett_2003}
{Jarrett}, T.~H., {Chester}, T., {Cutri}, R., {Schneider}, S.~E., \& {Huchra},
  J.~P. 2003, \aj, 125, 525

\bibitem[{{Je{\v{r}}{\'a}bkov{\'a}} {et~al.}(2018){Je{\v{r}}{\'a}bkov{\'a}},
  {Hasani Zonoozi}, {Kroupa}, {Beccari}, {Yan}, {Vazdekis}, \&
  {Zhang}}]{Jerabkova_2018}
{Je{\v{r}}{\'a}bkov{\'a}}, T., {Hasani Zonoozi}, A., {Kroupa}, P., {et~al.}
  2018, \aap, 620, A39

\bibitem[{{Joncour} {et~al.}(2018){Joncour}, {Duch{\^e}ne}, {Moraux}, \&
  {Motte}}]{Joncour_2018}
{Joncour}, I., {Duch{\^e}ne}, G., {Moraux}, E., \& {Motte}, F. 2018, \aap, 620,
  A27

\bibitem[{{Karachentsev} \& {Kaisina}(2013)}]{Karachentsev_2013_SFRproperties}
{Karachentsev}, I.~D. \& {Kaisina}, E.~I. 2013, \aj, 146, 46

\bibitem[{{Karachentsev} {et~al.}(2004){Karachentsev}, {Karachentseva},
  {Huchtmeier}, \& {Makarov}}]{Karachentsev_2004}
{Karachentsev}, I.~D., {Karachentseva}, V.~E., {Huchtmeier}, W.~K., \&
  {Makarov}, D.~I. 2004, \aj, 127, 2031

\bibitem[{{Karachentsev} {et~al.}(2013){Karachentsev}, {Makarov}, \&
  {Kaisina}}]{Karachentsev_2013}
{Karachentsev}, I.~D., {Makarov}, D.~I., \& {Kaisina}, E.~I. 2013, \aj, 145,
  101

\bibitem[{{Keller} \& {Kruijssen}(2022)}]{2022MNRAS.512..199K}
{Keller}, B.~W. \& {Kruijssen}, J.~M.~D. 2022, \mnras, 512, 199

\bibitem[{{Kennicutt}(1983)}]{Kennicutt_1983}
{Kennicutt}, R.~C., J. 1983, \apj, 272, 54

\bibitem[{{Kennicutt}(1998)}]{Kennicutt_1998}
{Kennicutt}, Robert~C., J. 1998, \araa, 36, 189

\bibitem[{{Kennicutt} \& {Evans}(2012)}]{Kennicutt_2012}
{Kennicutt}, R.~C. \& {Evans}, N.~J. 2012, \araa, 50, 531

\bibitem[{{Kewley} {et~al.}(2002){Kewley}, {Geller}, {Jansen}, \&
  {Dopita}}]{Kewley_2002}
{Kewley}, L.~J., {Geller}, M.~J., {Jansen}, R.~A., \& {Dopita}, M.~A. 2002,
  \aj, 124, 3135

\bibitem[{{Kirby} {et~al.}(2013){Kirby}, {Cohen}, {Guhathakurta}, {Cheng},
  {Bullock}, \& {Gallazzi}}]{Kirby_2013}
{Kirby}, E.~N., {Cohen}, J.~G., {Guhathakurta}, P., {et~al.} 2013, \apj, 779,
  102

\bibitem[{{Kirkpatrick} {et~al.}(2024){Kirkpatrick}, {Marocco}, {Gelino},
  {Raghu}, {Faherty}, {Bardalez Gagliuffi}, {Schurr}, {Apps}, {Schneider},
  {Meisner}, {Kuchner}, {Caselden}, {Smart}, {Casewell}, {Raddi}, {Kesseli},
  {Stevnbak Andersen}, {Antonini}, {Beaulieu}, {Bickle}, {Bilsing}, {Chieng},
  {Colin}, {Deen}, {Dereveanco}, {Doll}, {Durantini Luca}, {Frazer}, {Gantier},
  {Gramaize}, {Grant}, {Hamlet}, {Higashimura}, {Hyogo}, {Ja{\l}owiczor},
  {Jonkeren}, {Kabatnik}, {Kiwy}, {Martin}, {Michaels}, {Pendrill}, {Pessanha
  Machado}, {Pumphrey}, {Rothermich}, {Russwurm}, {Sainio}, {Sanchez},
  {Sapelkin-Tambling}, {Sch{\"u}mann}, {Selg-Mann}, {Singh}, {Stenner}, {Sun},
  {Tanner}, {Th{\'e}venot}, {Ventura}, {Voloshin}, {Walla}, {W{\k{e}}dracki},
  {Adorno}, {Aganze}, {Allers}, {Brooks}, {Burgasser}, {Calamari}, {Connor},
  {Costa}, {Eisenhardt}, {Gagn{\'e}}, {Gerasimov}, {Gonzales}, {Hsu}, {Kiman},
  {Li}, {Low}, {Mamajek}, {Pantoja}, {Popinchalk}, {Rees}, {Stern},
  {Su{\'a}rez}, {Theissen}, {Tsai}, {Vos}, {Zurek}, \& {The Backyard Worlds:
  Planet 9 Collaboration}}]{Kirkpatrick_2023}
{Kirkpatrick}, J.~D., {Marocco}, F., {Gelino}, C.~R., {et~al.} 2024, \apjs,
  271, 55

\bibitem[{{K{\"o}ppen} {et~al.}(2007){K{\"o}ppen}, {Weidner}, \&
  {Kroupa}}]{Koeppen_2007}
{K{\"o}ppen}, J., {Weidner}, C., \& {Kroupa}, P. 2007, \mnras, 375, 673

\bibitem[{{Kroupa}(1995)}]{Kroupa_1995}
{Kroupa}, P. 1995, \mnras, 277, 1507

\bibitem[{{Kroupa}(2001)}]{Kroupa_2001}
{Kroupa}, P. 2001, \mnras, 322, 231

\bibitem[{{Kroupa}(2002)}]{Kroupa_2002}
{Kroupa}, P. 2002, Science, 295, 82

\bibitem[{{Kroupa} {et~al.}(2023){Kroupa}, {Gjergo}, {Asencio}, {Haslbauer},
  {Pflamm-Altenburg}, {Wittenburg}, {Samaras}, {Thies}, \&
  {Oehm}}]{Kroupa_2023}
{Kroupa}, P., {Gjergo}, E., {Asencio}, E., {et~al.} 2023, in School and
  Workshops on Elementary Particle Physics and Gravity, 231

\bibitem[{{Kroupa} {et~al.}(2020{\natexlab{a}}){Kroupa}, {Haslbauer}, {Banik},
  {Nagesh}, \& {Pflamm-Altenburg}}]{Kroupa_2020}
{Kroupa}, P., {Haslbauer}, M., {Banik}, I., {Nagesh}, S.~T., \&
  {Pflamm-Altenburg}, J. 2020{\natexlab{a}}, \mnras, 497, 37

\bibitem[{{Kroupa} \& {Jerabkova}(2021)}]{Kroupa_Jerabkova_2021}
{Kroupa}, P. \& {Jerabkova}, T. 2021, arXiv e-prints, arXiv:2112.10788

\bibitem[{{Kroupa} {et~al.}(2020{\natexlab{b}}){Kroupa}, {Subr}, {Jerabkova},
  \& {Wang}}]{Kroupa_2020_SMBH}
{Kroupa}, P., {Subr}, L., {Jerabkova}, T., \& {Wang}, L. 2020{\natexlab{b}},
  \mnras, 498, 5652

\bibitem[{{Kroupa} \& {Weidner}(2003)}]{Kroupa_2003}
{Kroupa}, P. \& {Weidner}, C. 2003, \apj, 598, 1076

\bibitem[{{Kroupa} {et~al.}(2013){Kroupa}, {Weidner}, {Pflamm-Altenburg},
  {Thies}, {Dabringhausen}, {Marks}, \& {Maschberger}}]{Kroupa_2013}
{Kroupa}, P., {Weidner}, C., {Pflamm-Altenburg}, J., {et~al.} 2013, in Planets,
  Stars and Stellar Systems. Volume 5: Galactic Structure and Stellar
  Populations, ed. T.~D. {Oswalt} \& G.~{Gilmore}, Vol.~5, 115

\bibitem[{{Lada} \& {Lada}(2003)}]{Lada_2003}
{Lada}, C.~J. \& {Lada}, E.~A. 2003, \araa, 41, 57

\bibitem[{{Larsen}(2002)}]{2002AJ....124.1393L}
{Larsen}, S.~S. 2002, \aj, 124, 1393

\bibitem[{{Lee} \& {Yi}(2013)}]{2013ApJ...766...38L}
{Lee}, J. \& {Yi}, S.~K. 2013, \apj, 766, 38

\bibitem[{{Lee} {et~al.}(2009){Lee}, {Gil de Paz}, {Tremonti}, {Kennicutt},
  {Salim}, {Bothwell}, {Calzetti}, {Dalcanton}, {Dale}, {Engelbracht}, {Funes},
  {Johnson}, {Sakai}, {Skillman}, {van Zee}, {Walter}, \& {Weisz}}]{Lee_2009}
{Lee}, J.~C., {Gil de Paz}, A., {Tremonti}, C., {et~al.} 2009, \apj, 706, 599

\bibitem[{{Leja} {et~al.}(2015){Leja}, {van Dokkum}, {Franx}, \&
  {Whitaker}}]{Leja_2015}
{Leja}, J., {van Dokkum}, P.~G., {Franx}, M., \& {Whitaker}, K.~E. 2015, \apj,
  798, 115

\bibitem[{{Lelli}(2022)}]{Lelli_2022}
{Lelli}, F. 2022, Nature Astronomy, 6, 35

\bibitem[{{Lelli} {et~al.}(2014){Lelli}, {Verheijen}, \&
  {Fraternali}}]{Lelli_2014}
{Lelli}, F., {Verheijen}, M., \& {Fraternali}, F. 2014, \aap, 566, A71

\bibitem[{{Li} {et~al.}(2020){Li}, {Vogelsberger}, {Marinacci}, {Sales}, \&
  {Torrey}}]{Li_2020}
{Li}, H., {Vogelsberger}, M., {Marinacci}, F., {Sales}, L.~V., \& {Torrey}, P.
  2020, \mnras, 499, 5862

\bibitem[{{Li} {et~al.}(2023){Li}, {Liu}, {Zhang}, {Tian}, {Fu}, {Li}, \&
  {Yan}}]{Li_2023}
{Li}, J., {Liu}, C., {Zhang}, Z.-Y., {et~al.} 2023, \nat, 613, 460

\bibitem[{{Lieberz} \& {Kroupa}(2017)}]{Lieberz_2017}
{Lieberz}, P. \& {Kroupa}, P. 2017, \mnras, 465, 3775

\bibitem[{{Lu} {et~al.}(2023){Lu}, {Zhu}, {Cappellari}, {Li}, {Mao}, \&
  {Xu}}]{2023MNRAS.526.1022L}
{Lu}, S., {Zhu}, K., {Cappellari}, M., {et~al.} 2023, \mnras, 526, 1022

\bibitem[{{Ludlow} {et~al.}(2017){Ludlow}, {Ben{\'\i}tez-Llambay}, {Schaller},
  {Theuns}, {Frenk}, {Bower}, {Schaye}, {Crain}, {Navarro}, {Fattahi}, \&
  {Oman}}]{Ludlow_2017}
{Ludlow}, A.~D., {Ben{\'\i}tez-Llambay}, A., {Schaller}, M., {et~al.} 2017,
  Phys. Rev. Lett., 118, 161103

\bibitem[{{L{\"u}ghausen} {et~al.}(2015){L{\"u}ghausen}, {Famaey}, \&
  {Kroupa}}]{Lueghausen_2015}
{L{\"u}ghausen}, F., {Famaey}, B., \& {Kroupa}, P. 2015, Canadian Journal of
  Physics, 93, 232

\bibitem[{{Madau} \& {Dickinson}(2014)}]{Madau_Dickinson_2014}
{Madau}, P. \& {Dickinson}, M. 2014, \araa, 52, 415

\bibitem[{{Mahajan} {et~al.}(2019){Mahajan}, {Ashby}, {Willner}, {Barmby},
  {Fazio}, {Maragkoudakis}, {Raychaudhury}, \& {Zezas}}]{Mahajan_2019}
{Mahajan}, S., {Ashby}, M.~L.~N., {Willner}, S.~P., {et~al.} 2019, \mnras, 482,
  560

\bibitem[{{Ma{\'\i}z Apell{\'a}niz}(2006)}]{MaizApellaniz_2006}
{Ma{\'\i}z Apell{\'a}niz}, J. 2006, \aj, 131, 1184

\bibitem[{{Marasco} {et~al.}(2023){Marasco}, {Belfiore}, {Cresci}, {Lelli},
  {Venturi}, {Hunt}, {Concas}, {Marconi}, {Mannucci}, {Mingozzi}, {McLeod},
  {Kumari}, {Carniani}, {Vanzi}, \& {Ginolfi}}]{Marasco_2023}
{Marasco}, A., {Belfiore}, F., {Cresci}, G., {et~al.} 2023, \aap, 670, A92

\bibitem[{{Maraston}(2005)}]{Maraston_2005}
{Maraston}, C. 2005, \mnras, 362, 799

\bibitem[{{Marigo} {et~al.}(2013){Marigo}, {Bressan}, {Nanni}, {Girardi}, \&
  {Pumo}}]{Marigo_2013}
{Marigo}, P., {Bressan}, A., {Nanni}, A., {Girardi}, L., \& {Pumo}, M.~L. 2013,
  \mnras, 434, 488

\bibitem[{{Marks} {et~al.}(2012){Marks}, {Kroupa}, {Dabringhausen}, \&
  {Pawlowski}}]{Marks_2012}
{Marks}, M., {Kroupa}, P., {Dabringhausen}, J., \& {Pawlowski}, M.~S. 2012,
  \mnras, 422, 2246

\bibitem[{{Marks} {et~al.}(2014){Marks}, {Kroupa}, {Dabringhausen}, \&
  {Pawlowski}}]{Marks_2014}
{Marks}, M., {Kroupa}, P., {Dabringhausen}, J., \& {Pawlowski}, M.~S. 2014,
  \mnras, 442, 3315

\bibitem[{{Mart{\'\i}n-Navarro} {et~al.}(2019){Mart{\'\i}n-Navarro},
  {Lyubenova}, {van de Ven}, {Falc{\'o}n-Barroso}, {Coccato}, {Corsini},
  {Gadotti}, {Iodice}, {La Barbera}, {McDermid}, {Pinna}, {Sarzi}, {Viaene},
  {de Zeeuw}, \& {Zhu}}]{MartinNavarro_2019}
{Mart{\'\i}n-Navarro}, I., {Lyubenova}, M., {van de Ven}, G., {et~al.} 2019,
  \aap, 626, A124

\bibitem[{{Mart{\'\i}n-Navarro} {et~al.}(2015){Mart{\'\i}n-Navarro},
  {Vazdekis}, {La Barbera}, {Falc{\'o}n-Barroso}, {Lyubenova}, {van de Ven},
  {Ferreras}, {S{\'a}nchez}, {Trager}, {Garc{\'\i}a-Benito}, {Mast}, {Mendoza},
  {S{\'a}nchez-Bl{\'a}zquez}, {Gonz{\'a}lez Delgado}, {Walcher}, \& {CALIFA
  Team}}]{2015ApJ...806L..31M}
{Mart{\'\i}n-Navarro}, I., {Vazdekis}, A., {La Barbera}, F., {et~al.} 2015,
  \apjl, 806, L31

\bibitem[{{Matteucci}(1994)}]{Matteucci_1994}
{Matteucci}, F. 1994, \aap, 288, 57

\bibitem[{{McGaugh} \& {Schombert}(2014)}]{McGaugh_2014}
{McGaugh}, S.~S. \& {Schombert}, J.~M. 2014, \aj, 148, 77

\bibitem[{{McGaugh} {et~al.}(2017){McGaugh}, {Schombert}, \&
  {Lelli}}]{McGaugh_2017}
{McGaugh}, S.~S., {Schombert}, J.~M., \& {Lelli}, F. 2017, \apj, 851, 22

\bibitem[{{Meurer} {et~al.}(2009){Meurer}, {Wong}, {Kim}, {Hanish}, {Heckman},
  {Werk}, {Bland-Hawthorn}, {Dopita}, {Zwaan}, {Koribalski}, {Seibert},
  {Thilker}, {Ferguson}, {Webster}, {Putman}, {Knezek}, {Doyle}, {Drinkwater},
  {Hoopes}, {Kilborn}, {Meyer}, {Ryan-Weber}, {Smith}, \&
  {Staveley-Smith}}]{Meurer_2009}
{Meurer}, G.~R., {Wong}, O.~I., {Kim}, J.~H., {et~al.} 2009, \apj, 695, 765

\bibitem[{{Milgrom}(1983)}]{Milgrom_1983a}
{Milgrom}, M. 1983, \apj, 270, 365

\bibitem[{{Mitchell} {et~al.}(2014){Mitchell}, {Lacey}, {Cole}, \&
  {Baugh}}]{2014MNRAS.444.2637M}
{Mitchell}, P.~D., {Lacey}, C.~G., {Cole}, S., \& {Baugh}, C.~M. 2014, \mnras,
  444, 2637

\bibitem[{{Moustakas} {et~al.}(2006){Moustakas}, {Kennicutt}, \&
  {Tremonti}}]{Moustakas_2006}
{Moustakas}, J., {Kennicutt}, Robert~C., J., \& {Tremonti}, C.~A. 2006, \apj,
  642, 775

\bibitem[{{Mucciarelli} {et~al.}(2021){Mucciarelli}, {Massari}, {Minelli},
  {Romano}, {Bellazzini}, {Ferraro}, {Matteucci}, \&
  {Origlia}}]{Mucciarelli_2021}
{Mucciarelli}, A., {Massari}, D., {Minelli}, A., {et~al.} 2021, Nature
  Astronomy, 5, 1247

\bibitem[{{Nagesh} {et~al.}(2021){Nagesh}, {Banik}, {Thies}, {Kroupa},
  {Famaey}, {Wittenburg}, {Parziale}, \& {Haslbauer}}]{Nagesh_2021}
{Nagesh}, S.~T., {Banik}, I., {Thies}, I., {et~al.} 2021, Canadian Journal of
  Physics, 99, 607

\bibitem[{{Nagesh} {et~al.}(2023){Nagesh}, {Kroupa}, {Banik}, {Famaey},
  {Ghafourian}, {Roshan}, {Thies}, {Zhao}, \& {Wittenburg}}]{Nagesh_2023}
{Nagesh}, S.~T., {Kroupa}, P., {Banik}, I., {et~al.} 2023, \mnras, 519, 5128

\bibitem[{{Nelson} {et~al.}(2016){Nelson}, {van Dokkum}, {Momcheva}, {Brammer},
  {Wuyts}, {Franx}, {F{\"o}rster Schreiber}, {Whitaker}, \&
  {Skelton}}]{2016ApJ...817L...9N}
{Nelson}, E.~J., {van Dokkum}, P.~G., {Momcheva}, I.~G., {et~al.} 2016, \apjl,
  817, L9

\bibitem[{{Papadopoulos}(2010)}]{Papadopoulos_2010}
{Papadopoulos}, P.~P. 2010, \apj, 720, 226

\bibitem[{{Parikh} {et~al.}(2018){Parikh}, {Thomas}, {Maraston}, {Westfall},
  {Goddard}, {Lian}, {Meneses-Goytia}, {Jones}, {Vaughan}, {Andrews},
  {Bershady}, {Bizyaev}, {Brinkmann}, {Brownstein}, {Bundy}, {Drory},
  {Emsellem}, {Law}, {Newman}, {Roman-Lopes}, {Wake}, {Yan}, \&
  {Zheng}}]{Parikh_2018}
{Parikh}, T., {Thomas}, D., {Maraston}, C., {et~al.} 2018, \mnras, 477, 3954

\bibitem[{{Pastorelli} {et~al.}(2020){Pastorelli}, {Marigo}, {Girardi},
  {Aringer}, {Chen}, {Rubele}, {Trabucchi}, {Bladh}, {Boyer}, {Bressan},
  {Dalcanton}, {Groenewegen}, {Lebzelter}, {Mowlavi}, {Chubb}, {Cioni}, {de
  Grijs}, {Ivanov}, {Nanni}, {van Loon}, \& {Zaggia}}]{Pastorelli_2020}
{Pastorelli}, G., {Marigo}, P., {Girardi}, L., {et~al.} 2020, \mnras, 498, 3283

\bibitem[{{Pastorelli} {et~al.}(2019){Pastorelli}, {Marigo}, {Girardi}, {Chen},
  {Rubele}, {Trabucchi}, {Aringer}, {Bladh}, {Bressan}, {Montalb{\'a}n},
  {Boyer}, {Dalcanton}, {Eriksson}, {Groenewegen}, {H{\"o}fner}, {Lebzelter},
  {Nanni}, {Rosenfield}, {Wood}, \& {Cioni}}]{Pastorelli_2019}
{Pastorelli}, G., {Marigo}, P., {Girardi}, L., {et~al.} 2019, \mnras, 485, 5666

\bibitem[{{Pflamm-Altenburg} \& {Kroupa}(2008)}]{PflammAltenburg_2008}
{Pflamm-Altenburg}, J. \& {Kroupa}, P. 2008, \nat, 455, 641

\bibitem[{{Pflamm-Altenburg} \& {Kroupa}(2009)}]{PflammAltenburg_2009b}
{Pflamm-Altenburg}, J. \& {Kroupa}, P. 2009, \apj, 706, 516

\bibitem[{{Pflamm-Altenburg} {et~al.}(2007){Pflamm-Altenburg}, {Weidner}, \&
  {Kroupa}}]{PflammAltenburg_2007}
{Pflamm-Altenburg}, J., {Weidner}, C., \& {Kroupa}, P. 2007, \apj, 671, 1550

\bibitem[{{Pflamm-Altenburg} {et~al.}(2009){Pflamm-Altenburg}, {Weidner}, \&
  {Kroupa}}]{PflammAltenburg_2009a}
{Pflamm-Altenburg}, J., {Weidner}, C., \& {Kroupa}, P. 2009, \mnras, 395, 394

\bibitem[{{Planck Collaboration XIII}(2016)}]{Planck_2016_IllustrisTNG}
{Planck Collaboration XIII}. 2016, \aap, 594, A13

\bibitem[{{Ploeckinger} {et~al.}(2014){Ploeckinger}, {Hensler}, {Recchi},
  {Mitchell}, \& {Kroupa}}]{Ploeckinger_2014}
{Ploeckinger}, S., {Hensler}, G., {Recchi}, S., {Mitchell}, N., \& {Kroupa}, P.
  2014, \mnras, 437, 3980

\bibitem[{{Ploeckinger} {et~al.}(2015){Ploeckinger}, {Recchi}, {Hensler}, \&
  {Kroupa}}]{Ploeckinger_2015}
{Ploeckinger}, S., {Recchi}, S., {Hensler}, G., \& {Kroupa}, P. 2015, \mnras,
  447, 2512

\bibitem[{{Rasmussen Cueto} {et~al.}(2023){Rasmussen Cueto}, {Hutter}, {Dayal},
  {Gottl{\"o}ber}, {Heintz}, {Mason}, {Trebitsch}, \&
  {Yepes}}]{2023arXiv231212109R}
{Rasmussen Cueto}, E., {Hutter}, A., {Dayal}, P., {et~al.} 2023, arXiv
  e-prints, arXiv:2312.12109

\bibitem[{{Recchi} {et~al.}(2009){Recchi}, {Calura}, \& {Kroupa}}]{Recchi_2009}
{Recchi}, S., {Calura}, F., \& {Kroupa}, P. 2009, \aap, 499, 711

\bibitem[{{Reid} {et~al.}(2002){Reid}, {Gizis}, \& {Hawley}}]{Reid_2002}
{Reid}, I.~N., {Gizis}, J.~E., \& {Hawley}, S.~L. 2002, \aj, 124, 2721

\bibitem[{{Ren} {et~al.}(2024){Ren}, {Jiang}, {Wang}, {Yang}, \&
  {Yan}}]{Ren_2024}
{Ren}, Y., {Jiang}, B., {Wang}, Y., {Yang}, M., \& {Yan}, Z. 2024, \apj, 966,
  25

\bibitem[{{Rieke} {et~al.}(2009){Rieke}, {Alonso-Herrero}, {Weiner},
  {P{\'e}rez-Gonz{\'a}lez}, {Blaylock}, {Donley}, \& {Marcillac}}]{Rieke_2009}
{Rieke}, G.~H., {Alonso-Herrero}, A., {Weiner}, B.~J., {et~al.} 2009, \apj,
  692, 556

\bibitem[{{Rosenfield} {et~al.}(2016){Rosenfield}, {Marigo}, {Girardi},
  {Dalcanton}, {Bressan}, {Williams}, \& {Dolphin}}]{Rosenfield_2016}
{Rosenfield}, P., {Marigo}, P., {Girardi}, L., {et~al.} 2016, \apj, 822, 73

\bibitem[{{Salpeter}(1955)}]{Salpeter_1955}
{Salpeter}, E.~E. 1955, \apj, 121, 161

\bibitem[{{Schaye} {et~al.}(2015){Schaye}, {Crain}, {Bower}, {Furlong},
  {Schaller}, {Theuns}, {Dalla Vecchia}, {Frenk}, {McCarthy}, {Helly},
  {Jenkins}, {Rosas-Guevara}, {White}, {Baes}, {Booth}, {Camps}, {Navarro},
  {Qu}, {Rahmati}, {Sawala}, {Thomas}, \& {Trayford}}]{2015MNRAS.446..521S}
{Schaye}, J., {Crain}, R.~A., {Bower}, R.~G., {et~al.} 2015, \mnras, 446, 521

\bibitem[{{Schneider} {et~al.}(2018){Schneider}, {Sana}, {Evans},
  {Bestenlehner}, {Castro}, {Fossati}, {Gr{\"a}fener}, {Langer},
  {Ram{\'\i}rez-Agudelo}, {Sab{\'\i}n-Sanjuli{\'a}n}, {Sim{\'o}n-D{\'\i}az},
  {Tramper}, {Crowther}, {de Koter}, {de Mink}, {Dufton}, {Garcia}, {Gieles},
  {H{\'e}nault-Brunet}, {Herrero}, {Izzard}, {Kalari}, {Lennon}, {Ma{\'\i}z
  Apell{\'a}niz}, {Markova}, {Najarro}, {Podsiadlowski}, {Puls}, {Taylor}, {van
  Loon}, {Vink}, \& {Norman}}]{Schneider_2018}
{Schneider}, F.~R.~N., {Sana}, H., {Evans}, C.~J., {et~al.} 2018, Science, 359,
  69

\bibitem[{{Schombert} {et~al.}(2019){Schombert}, {McGaugh}, \&
  {Lelli}}]{Schombert_2019}
{Schombert}, J., {McGaugh}, S., \& {Lelli}, F. 2019, \mnras, 483, 1496

\bibitem[{{Schulz} {et~al.}(2015){Schulz}, {Pflamm-Altenburg}, \&
  {Kroupa}}]{2015A&A...582A..93S}
{Schulz}, C., {Pflamm-Altenburg}, J., \& {Kroupa}, P. 2015, \aap, 582, A93

\bibitem[{{Sharda} \& {Krumholz}(2022)}]{2022MNRAS.509.1959S}
{Sharda}, P. \& {Krumholz}, M.~R. 2022, \mnras, 509, 1959

\bibitem[{{Smith}(2020)}]{Smith_2020}
{Smith}, R.~J. 2020, \araa, 58, 577

\bibitem[{{Sneppen} {et~al.}(2022){Sneppen}, {Steinhardt}, {Hensley}, {Jermyn},
  {Mostafa}, \& {Weaver}}]{Sneppen_2022}
{Sneppen}, A., {Steinhardt}, C.~L., {Hensley}, H., {et~al.} 2022, \apj, 931, 57

\bibitem[{{Speagle} {et~al.}(2014){Speagle}, {Steinhardt}, {Capak}, \&
  {Silverman}}]{Speagle_2014}
{Speagle}, J.~S., {Steinhardt}, C.~L., {Capak}, P.~L., \& {Silverman}, J.~D.
  2014, \apjs, 214, 15

\bibitem[{{Tomczak} {et~al.}(2016){Tomczak}, {Quadri}, {Tran}, {Labb{\'e}},
  {Straatman}, {Papovich}, {Glazebrook}, {Allen}, {Brammer}, {Cowley},
  {Dickinson}, {Elbaz}, {Inami}, {Kacprzak}, {Morrison}, {Nanayakkara},
  {Persson}, {Rees}, {Salmon}, {Schreiber}, {Spitler}, \&
  {Whitaker}}]{Tomczak_2016}
{Tomczak}, A.~R., {Quadri}, R.~F., {Tran}, K.-V.~H., {et~al.} 2016, \apj, 817,
  118

\bibitem[{{Trabucchi} {et~al.}(2021){Trabucchi}, {Wood}, {Mowlavi},
  {Pastorelli}, {Marigo}, {Girardi}, \& {Lebzelter}}]{Trabucchi_2021}
{Trabucchi}, M., {Wood}, P.~R., {Mowlavi}, N., {et~al.} 2021, \mnras, 500, 1575

\bibitem[{{Valentino} {et~al.}(2021){Valentino}, {Daddi}, {Puglisi}, {Magdis},
  {Kokorev}, {Liu}, {Madden}, {G{\'o}mez-Guijarro}, {Lee}, {Cortzen},
  {Circosta}, {Delvecchio}, {Mullaney}, {Gao}, {Gobat}, {Aravena}, {Jin},
  {Fujimoto}, {Silverman}, \& {Dannerbauer}}]{Valentino_2021}
{Valentino}, F., {Daddi}, E., {Puglisi}, A., {et~al.} 2021, \aap, 654, A165

\bibitem[{{van Dokkum} \& {Conroy}(2021)}]{vanDokkum_2021}
{van Dokkum}, P. \& {Conroy}, C. 2021, \apj, 923, 43

\bibitem[{{Vazdekis} {et~al.}(1997){Vazdekis}, {Peletier}, {Beckman}, \&
  {Casuso}}]{Vazdekis_1997}
{Vazdekis}, A., {Peletier}, R.~F., {Beckman}, J.~E., \& {Casuso}, E. 1997,
  \apjs, 111, 203

\bibitem[{{Wang} {et~al.}(2024){Wang}, {Leja}, {Atek}, {Labb{\'e}}, {Li},
  {Bezanson}, {Brammer}, {Cutler}, {Dayal}, {Furtak}, {Greene}, {Kokorev},
  {Pan}, {Price}, {Suess}, {Weaver}, {Whitaker}, \&
  {Williams}}]{2023arXiv231006781W}
{Wang}, B., {Leja}, J., {Atek}, H., {et~al.} 2024, \apj, 963, 74

\bibitem[{{Watts} {et~al.}(2018){Watts}, {Meurer}, {Lagos}, {Bruzzese},
  {Kroupa}, \& {Jerabkova}}]{Watts_2018}
{Watts}, A.~B., {Meurer}, G.~R., {Lagos}, C. D.~P., {et~al.} 2018, \mnras, 477,
  5554

\bibitem[{{Weidner} {et~al.}(2013{\natexlab{a}}){Weidner}, {Ferreras},
  {Vazdekis}, \& {La Barbera}}]{2013MNRAS.435.2274W}
{Weidner}, C., {Ferreras}, I., {Vazdekis}, A., \& {La Barbera}, F.
  2013{\natexlab{a}}, \mnras, 435, 2274

\bibitem[{{Weidner} \& {Kroupa}(2006)}]{Weidner_2006}
{Weidner}, C. \& {Kroupa}, P. 2006, \mnras, 365, 1333

\bibitem[{{Weidner} {et~al.}(2004){Weidner}, {Kroupa}, \&
  {Larsen}}]{Weidner_2004b}
{Weidner}, C., {Kroupa}, P., \& {Larsen}, S.~S. 2004, \mnras, 350, 1503

\bibitem[{{Weidner} {et~al.}(2011){Weidner}, {Kroupa}, \&
  {Pflamm-Altenburg}}]{2011MNRAS.412..979W}
{Weidner}, C., {Kroupa}, P., \& {Pflamm-Altenburg}, J. 2011, \mnras, 412, 979

\bibitem[{{Weidner} {et~al.}(2013{\natexlab{b}}){Weidner}, {Kroupa}, \&
  {Pflamm-Altenburg}}]{2013MNRAS.434...84W}
{Weidner}, C., {Kroupa}, P., \& {Pflamm-Altenburg}, J. 2013{\natexlab{b}},
  \mnras, 434, 84

\bibitem[{{Weidner} {et~al.}(2014){Weidner}, {Kroupa}, \&
  {Pflamm-Altenburg}}]{2014MNRAS.441.3348W}
{Weidner}, C., {Kroupa}, P., \& {Pflamm-Altenburg}, J. 2014, \mnras, 441, 3348

\bibitem[{{Weidner} {et~al.}(2013{\natexlab{c}}){Weidner}, {Kroupa},
  {Pflamm-Altenburg}, \& {Vazdekis}}]{2013MNRAS.436.3309W}
{Weidner}, C., {Kroupa}, P., {Pflamm-Altenburg}, J., \& {Vazdekis}, A.
  2013{\natexlab{c}}, \mnras, 436, 3309

\bibitem[{{Wilkins} {et~al.}(2008){Wilkins}, {Trentham}, \&
  {Hopkins}}]{Wilkins_2008}
{Wilkins}, S.~M., {Trentham}, N., \& {Hopkins}, A.~M. 2008, \mnras, 385, 687

\bibitem[{{Willmer}(2018)}]{Willmer_2018}
{Willmer}, C. N.~A. 2018, \apjs, 236, 47

\bibitem[{{Wittenburg} {et~al.}(2020){Wittenburg}, {Kroupa}, \&
  {Famaey}}]{Wittenburg_2020}
{Wittenburg}, N., {Kroupa}, P., \& {Famaey}, B. 2020, \apj, 890, 173

\bibitem[{{Yan} {et~al.}(2017){Yan}, {Jerabkova}, \& {Kroupa}}]{Yan_2017}
{Yan}, Z., {Jerabkova}, T., \& {Kroupa}, P. 2017, \aap, 607, A126

\bibitem[{{Yan} {et~al.}(2019{\natexlab{a}}){Yan}, {Jerabkova}, \&
  {Kroupa}}]{Yan_2019b}
{Yan}, Z., {Jerabkova}, T., \& {Kroupa}, P. 2019{\natexlab{a}}, \aap, 632, A110

\bibitem[{{Yan} {et~al.}(2020){Yan}, {Jerabkova}, \& {Kroupa}}]{Yan_2020}
{Yan}, Z., {Jerabkova}, T., \& {Kroupa}, P. 2020, \aap, 637, A68

\bibitem[{{Yan} {et~al.}(2023){Yan}, {Jerabkova}, \& {Kroupa}}]{Yan_2023}
{Yan}, Z., {Jerabkova}, T., \& {Kroupa}, P. 2023, \aap, 670, A151

\bibitem[{{Yan} {et~al.}(2019{\natexlab{b}}){Yan}, {Jerabkova}, {Kroupa}, \&
  {Vazdekis}}]{Yan_2019a}
{Yan}, Z., {Jerabkova}, T., {Kroupa}, P., \& {Vazdekis}, A. 2019{\natexlab{b}},
  \aap, 629, A93

\bibitem[{{Yan} {et~al.}(2021){Yan}, {Je{\v{r}}{\'a}bkov{\'a}}, \&
  {Kroupa}}]{Yan_2021}
{Yan}, Z., {Je{\v{r}}{\'a}bkov{\'a}}, T., \& {Kroupa}, P. 2021, \aap, 655, A19

\bibitem[{{Yan} {et~al.}(2024){Yan}, {Li}, {Kroupa}, {Jerabkova}, {Gjergo}, \&
  {Zhang}}]{Yan24}
{Yan}, Z., {Li}, J., {Kroupa}, P., {et~al.} 2024, ApJ accepted

\bibitem[{{Yu} \& {Wang}(2016)}]{Yu_2016}
{Yu}, H. \& {Wang}, F.~Y. 2016, \apj, 820, 114

\bibitem[{{Yun} {et~al.}(2001){Yun}, {Reddy}, \& {Condon}}]{Yun_2001}
{Yun}, M.~S., {Reddy}, N.~A., \& {Condon}, J.~J. 2001, \apj, 554, 803

\bibitem[{{Zezas} \& {Buat}(2021)}]{Zezas_2021}
{Zezas}, A. \& {Buat}, V. 2021, {Star Formation Rates in Galaxies}, Vol.~55

\bibitem[{{Zhang} {et~al.}(2018){Zhang}, {Romano}, {Ivison}, {Papadopoulos}, \&
  {Matteucci}}]{Zhang_2018}
{Zhang}, Z.-Y., {Romano}, D., {Ivison}, R.~J., {Papadopoulos}, P.~P., \&
  {Matteucci}, F. 2018, \nat, 558, 260

\bibitem[{{Zhou} {et~al.}(2019){Zhou}, {Mo}, {Li}, {Zheng}, {Li}, {Du}, {Mao},
  {Parikh}, {Lane}, \& {Thomas}}]{Zhou_2019}
{Zhou}, S., {Mo}, H.~J., {Li}, C., {et~al.} 2019, \mnras, 485, 5256

\bibitem[{{Zhu} {et~al.}(2022){Zhu}, {van de Ven}, {Leaman}, {Pillepich},
  {Coccato}, {Ding}, {Falc{\'o}n-Barroso}, {Iodice}, {Navarro}, {Pinna},
  {Corsini}, {Gadotti}, {Fahrion}, {Lyubenova}, {Mao}, {McDermid}, {Poci},
  {Sarzi}, \& {de Zeeuw}}]{2022A&A...664A.115Z}
{Zhu}, L., {van de Ven}, G., {Leaman}, R., {et~al.} 2022, \aap, 664, A115

\bibitem[{{Zonoozi} {et~al.}(2019){Zonoozi}, {Mahani}, \&
  {Kroupa}}]{Zonoozi_2019}
{Zonoozi}, A.~H., {Mahani}, H., \& {Kroupa}, P. 2019, \mnras, 483, 46

\end{thebibliography}

\begin{appendix}

\section{M/L evolution for stellar populations with fixed metallicities} \label{appendix_MLR_canonicalIMF}

\begin{figure*}
    \includegraphics[width=\linewidth]{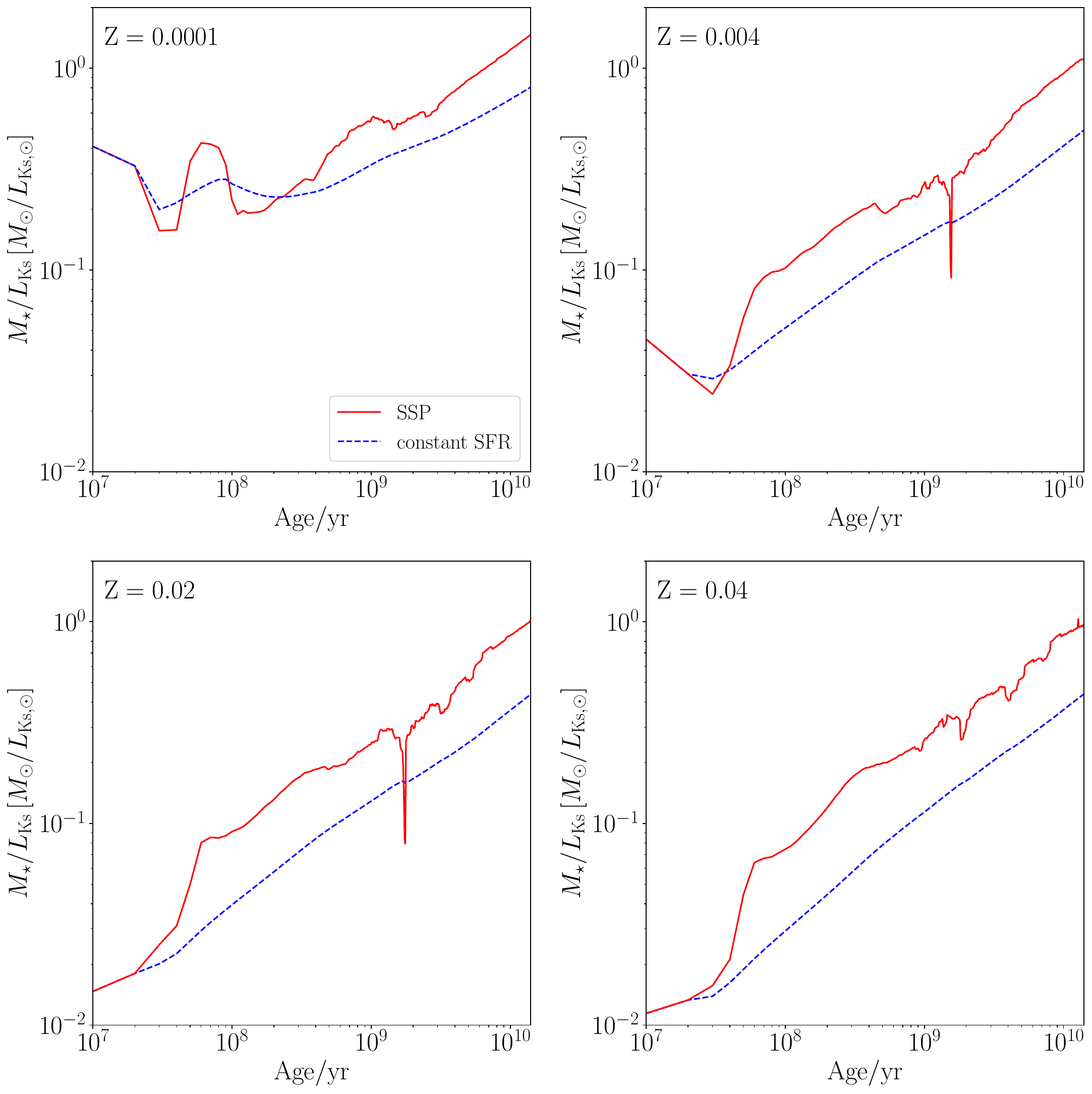}
    \caption{Time evolution of the Ks-band mass-to-light ratio, $M_\star/L_{\mathrm{Ks}}$, for stellar populations with a fixed canonical \citep{Kroupa_2001} IMF and different metallicities denoted in the upper left corner of each panel. Red solid curves are SSPs formed at time zero.
    The sharp drop in the SSP models of $M_\star/L_{\mathrm{Ks}}$ at $t \approx 2\times10^{9}$~yr is caused by a significant increase in the number of AGB stars. The blue dashed curves are stellar populations with the same fixed metallicity but a constant SFR over time, demonstrating a smooth curve of lower $M_\star/L_{\mathrm{Ks}}$ value due to the luminosity contribution from young stars. The exact mass of the SSP or the SFR for the constant SFR model does not affect the IMF by definition, and therefore, does not affect the $M_\star/L_{\mathrm{Ks}}$ evolution curve.} \label{figure_evolution_MLR_canonicalIMF}
\end{figure*}

As an example, Fig.~\ref{figure_evolution_MLR_canonicalIMF} shows the time evolution of $M_\star/L_{\mathrm{Ks}}$ for single stellar populations (SSPs) and
constant SFR models with different metallicities ([Z] = -2.17, -0.57, 0.13, 0.43) assuming the canonical IMF \citep{Kroupa_2001}. The sharp decrease at $\approx 2\times 10^9$~yr is due to a pronounced peak in the production rate of AGB stars as explained in \citet{Girardi_1998} and \citet{Girardi_2013}. Our result is qualitatively in agreement with previous studies (e.g. \citealt{Maraston_2005} their figure~23 and \citealt{2015A&A...575A.128B} their figure~10). Other than the choices of wavelength, metallicity, and time resolution, the differences in the shape of the curve and the exact time of the pronounced luminosity peak are also due to updated stellar and TP-AGB evolution models of the PARSEC code.

\section{Fitting parameters} \label{appendix_fittingparameters}
Table~\ref{tab:fitting_parameters} lists the polynomial coefficients for different metallicities of the fitting functions presented in Section~\ref{subsec: SFR-Halpha relation}. We note that these fifth-order polynomial functions are only valid for the fitted range of parameters (see table caption) and should not be extrapolated.

\begin{table*}
 \caption{\label{t7} \textbf{Polynomial coefficients of the fitting functions presented in Section~\ref{subsec: SFR-Halpha relation}.}}
  \centering
	\resizebox{\linewidth}{!}{\begin{tabular}{lllllllllll}
	\hline
    Relation & $Z$ &$\rm{[Z]}$ & $a$ & $b$ & $c$ & $d$ & $e$ & $f$ \\ \hline 
    $\psi$--$L_{\mathrm{H\alpha}}$ & $9.0 \times 10^{-5}$ & $-2.20$  & $2.2238\times 10^{-5}$ &  $1.2128 \times 10^{-4}$ &   $-1.5783 \times 10^{-3}$ &   $2.0957\times 10^{-2}$ &  $7.3060 \times 10^{-1}$ &   $-5.7930\times 10^{-1}$ \\ 
    
    (Equation~(\ref{eq:IGIMF_Halpha_SFR_fit})) & $9.6 \times 10^{-5}$ & $-2.17$  & $2.2364 \times 10^{-5}$ &  $1.2240 \times 10^{-4}$ &   $-1.5890 \times 10^{-3}$ &   $2.0895 \times 10^{-2}$ &  $7.3092 \times 10^{-1}$ &   $-5.7876 \times 10^{-1}$ \\ 

    & $0.00019$ & $-1.87$  & $2.2504  \times 10^{-5}$ &  $1.3788 \times 10^{-4}$ &  $-1.5479 \times 10^{-3}$ &   $2.0022 \times 10^{-2}$ &   $7.2795 \times 10^{-1}$ &   $-5.4425 \times 10^{-1}$\\ 
    
    & $0.00048$ & $-1.47$ & $1.8490 \times 10^{-5}$ &  $1.1702 \times 10^{-4}$ &  $-1.2605 \times 10^{-3}$ &   $2.0782 \times 10^{-2}$ &   $7.2005 \times 10^{-1}$ &   $-5.0532 \times 10^{-1}$ \\ 

    & $0.00096$ & $-1.17$ & $1.9246 \times 10^{-5}$ &  $1.3371 \times 10^{-4}$ &   $-1.2517 \times 10^{-3}$ &   $1.9954\times 10^{-2}$ &   $7.1826 \times 10^{-1}$ &   $-4.7476\times 10^{-1}$ \\
    
    & $0.0038$ & $-0.57$ & $1.5919 \times 10^{-5}$ &  $1.2457 \times 10^{-4}$ &   $-9.7997 \times 10^{-4}$ &   $2.0551 \times 10^{-2}$ &   $7.0787 \times 10^{-1}$ &   $-4.1389 \times 10^{-1}$ \\ 
    
    & $0.0078$ & $-0.26$ & $1.9228 \times 10^{-5}$ &  $1.8041\times 10^{-4}$ &   $-1.1073 \times 10^{-3}$ &   $1.8449 \times 10^{-2}$ &  $7.0538  \times 10^{-1}$ &   $-3.4245 \times 10^{-1}$ \\ 
   
    & $0.0096$ & $-0.17$ & $1.7296 \times 10^{-5}$ &  $1.6136 \times 10^{-4}$ &  $-9.4291 \times 10^{-4}$ &   $1.9413 \times 10^{-2}$ &   $7.0002 \times 10^{-1}$ &   $-3.2063 \times 10^{-1}$ \\ 
    
    & $0.0142$ & $0.0$ & $2.3876 \times 10^{-5}$ &  $2.4911 \times 10^{-4}$ &   $-1.2368 \times 10^{-3}$ &   $1.6128 \times 10^{-2}$ &   $7.0077 \times 10^{-1}$ &   $-2.5619 \times 10^{-1}$
    \\ 
   
    & $0.019$ & $0.13$ & $2.0135 \times 10^{-5}$ &  $2.3304 \times 10^{-4}$ &   $-8.7799 \times 10^{-4}$ &   $1.7239 \times 10^{-2}$ &   $6.8696 \times 10^{-1}$ &   $-2.0074 \times 10^{-1}$
    \\ 
    
    & $0.029$ & $0.31$ & $1.8947 \times 10^{-5}$ &  $2.4694 \times 10^{-4}$ &  $-5.3669\times 10^{-4}$ &   $1.8314\times 10^{-2}$ &   $6.6154 \times 10^{-1}$ &   $-4.9672 \times 10^{-2}$ \\ 
    
    & $0.038$ & $0.43$ & $2.1314\times 10^{-5}$ &  $2.7855\times 10^{-4}$ &   $-3.6256\times 10^{-4}$ &   $1.9320 \times 10^{-2}$ &   $6.3030 \times 10^{-1}$ &   $1.4218 \times 10^{-1}$ \\
    
    & $0.040$ & $0.45$ & $1.5181 \times 10^{-5}$ &  $1.9780 \times 10^{-4}$ &   $-6.5759 \times 10^{-5}$ &   $2.2975 \times 10^{-2}$ &   $6.2133 \times 10^{-1}$ &   $1.6575 \times 10^{-1}$   \\ \hline

    $\Phi-L_{\mathrm{H\alpha}}$ & $9.0 \times 10^{-5}$ & $-2.20$  & $2.2238 \times 10^{-5}$ &  $1.2128 \times 10^{-4}$ &   $-1.5783 \times 10^{-3}$ &   $2.0957 \times 10^{-2}$ &  $-2.6940 \times 10^{-1}$ &   $-4.7893 \times 10^{-1}$ \\ 

    (Equation~(\ref{eq:IGIMF_Kennicutt_SFR_fit}))& $9.6 \times 10^{-5}$ & $-2.17$ & $2.2364 \times 10^{-5}$ &  $1.2240 \times 10^{-4}$ &   $-1.5890 \times 10^{-3}$ &   $2.0895\times 10^{-2}$ &  $-2.6908 \times 10^{-1}$ &   $-4.7839\times 10^{-1}$ \\ 

    & $0.00019$ & $-1.87$ & $2.2504 \times 10^{-5}$ &  $1.3788\times 10^{-4}$ &  $-1.5479 \times 10^{-3}$ &   $2.0022 \times 10^{-2}$ &   $-2.7205 \times 10^{-1}$ &   $-4.4388 \times 10^{-1}$ \\ 
   
    & $0.00048$ & $-1.47$ & $1.8490 \times 10^{-5}$ &  $1.1702 \times 10^{-4}$ &   $-1.2605 \times 10^{-3}$ &   $2.0782 \times 10^{-2}$ &   $-2.7995 \times 10^{-1}$ &   $-4.0495 \times 10^{-1}$ \\

    & $0.00096$ & $-1.17$ & $1.9246 \times 10^{-5}$ &  $1.3371 \times 10^{-4}$ &   $-1.2517 \times 10^{-3}$ &   $1.9954 \times 10^{-2}$ &   $ -2.8174 \times 10^{-1}$ &   $-3.7439 \times 10^{-1}$ \\ 

    & $0.0038$ & $-0.57$ & $1.5919 \times 10^{-5}$ &  $1.2457\times 10^{-4}$ &   $-9.7997 \times 10^{-4}$ &   $2.0551 \times 10^{-2}$ &  $-2.9213 \times 10^{-1}$ &   $-3.1351 \times 10^{-1}$ \\ 
   
    & $0.0078$ & $-0.26$ & $1.9228 \times 10^{-5}$ &  $1.8041 \times 10^{-4}$ &  $-1.1073 \times 10^{-3}$ &   $1.8449\times 10^{-2}$ &   $-2.9462 \times 10^{-1}$ &   $ -2.4208\times 10^{-1}$ \\ 

    & $0.0096$ & $-0.17$ & $ 1.7296 \times 10^{-5}$ &  $1.6136 \times 10^{-4}$ &   $-9.4291 \times 10^{-4}$ &   $1.9413\times 10^{-2}$ &   $-2.9998 \times 10^{-1}$ &   $-2.2026 \times 10^{-1}$ \\

    & $0.0142$ & $0.0$ & $2.3876 \times 10^{-5}$ &  $2.4911 \times 10^{-4}$ &   $-1.2368 \times 10^{-3}$ &   $1.6128 \times 10^{-2}$ &   $ -2.9923 \times 10^{-1}$ &   $-1.5582 \times 10^{-1}$
    \\ 
   
    & $0.019$ & $0.13$ & $2.0135 \times 10^{-5}$ &  $2.3304 \times 10^{-4}$ &  $-8.7799\times 10^{-4}$ &   $1.7239\times 10^{-2}$ &   $-3.1304 \times 10^{-1}$ &   $-1.0037 \times 10^{-1}$ \\ 
  
    & $0.029$ & $0.31$ & $1.8947 \times 10^{-5}$ &  $2.4694\times 10^{-4}$ &   $-5.3669 \times 10^{-4}$ & $1.8314\times 10^{-2}$ &   $-3.3846 \times 10^{-1}$ &   $  5.0707\times 10^{-2}$ &  \\
    
    & $0.038$ & $0.43$ & $2.1314\times 10^{-5}$ & $2.7855\times 10^{-4}$ &  $-3.6256 \times 10^{-4}$ &   $ 1.9320\times 10^{-2}$ &   $-3.6970 \times 10^{-1}$ &   $2.4256 \times 10^{-1}$ \\
    
    & $0.040$ & $0.45$ & $1.5181 \times 10^{-5}$ &  $1.9780 \times 10^{-4}$ &   $-6.5760 \times 10^{-5}$ &   $2.2975\times 10^{-2}$ &   $-3.7867\times 10^{-1}$ &   $2.6612 \times 10^{-1}$ \\
    \hline
    $\psi_{\mathrm{K98}}-\psi_{\mathrm{IGIMF}}$ & $9.0 \times 10^{-5}$ & $-2.20$  & $2.2238 \times 10^{-5}$ &  $1.3244 \times 10^{-4}$ &   $-1.5273 \times 10^{-3}$ &   $2.0489 \times 10^{-2}$ &  $7.3476 \times 10^{-1}$ &   $-5.0576 \times 10^{-1}$ \\ 
    
    (Equation~(\ref{eq:IGIMF_Kennicutt_SFR_fit_2})) & $9.6 \times 10^{-5}$ & $-2.17$ & $2.2364\times 10^{-5}$ &  $1.3362 \times 10^{-4}$ &   $-1.5376\times 10^{-3}$ &   $2.0424 \times 10^{-2}$ &  $7.3506 \times 10^{-1}$ &   $-5.0519 \times 10^{-1}$ \\ 
    
    & $0.00019$ & $-1.87$ & $2.2504\times 10^{-5}$ &  $1.4917 \times 10^{-4}$ &  $-1.4903\times 10^{-3}$ &   $1.9564 \times 10^{-2}$ &   $7.3192 \times 10^{-1}$ &   $-4.7098 \times 10^{-1}$ \\ 
    
    & $0.00048$ & $-1.47$ & $1.8490\times 10^{-5}$ &  $1.2630\times 10^{-4}$ &   $-1.2116\times 10^{-3}$ &   $2.0410 \times 10^{-2}$ &   $7.2418 \times 10^{-1}$ &   $-4.3284 \times 10^{-1}$ \\
    
    & $0.00096$ & $-1.17$ & $1.9246\times 10^{-5}$ &  $1.4337\times 10^{-4}$ &   $-1.1961\times 10^{-3}$ &   $1.9585\times 10^{-2}$ &   $7.2223 \times 10^{-1}$ &   $-4.0246 \times 10^{-1}$ \\ 
   
    & $0.0038$ & $-0.57$ & $1.5919\times 10^{-5}$ &  $1.3256\times 10^{-4}$ &   $-9.2835\times 10^{-4}$ &   $2.0264\times 10^{-2}$ &  $7.1196\times 10^{-1}$ &   $-3.4263\times 10^{-1}$ \\ 
   
    & $0.0078$ & $-0.26$ & $1.9228\times 10^{-5}$ &  $1.9005 \times 10^{-4}$ &  $-1.0329\times 10^{-3}$ &   $1.8126\times 10^{-2}$ &   $7.0905\times 10^{-1}$ &   $-2.7146\times 10^{-1}$ \\ 
    
    & $0.0096$ & $-0.17$ & $1.7296\times 10^{-5}$ &  $1.7004\times 10^{-4}$ &   $-8.7638\times 10^{-4}$ &   $1.9139\times 10^{-2}$ &   $7.0389\times 10^{-1}$ &   $-2.5017\times 10^{-1}$ \\
   
    & $0.0142$ & $0.0$ & $2.3876 \times 10^{-5}$ &  $2.61089 \times 10^{-4}$ &   $-1.1344 \times 10^{-3}$ &   $1.5771 \times 10^{-2}$ &   $7.0397 \times 10^{-1}$ &   $-1.8569 \times 10^{-1}$
    \\ 
    
    & $0.019$ & $0.13$ & $2.0135\times 10^{-5}$ &  $2.4314\times 10^{-4}$ &  $-7.8240\times 10^{-4}$ &   $1.6989\times 10^{-2}$ &   $6.9040\times 10^{-1}$ &   $-1.3162\times 10^{-1}$ \\ 
    
    & $0.029$ & $0.31$ & $1.8947\times 10^{-5}$ &  $2.5645\times 10^{-4}$ &   $-4.3564\times 10^{-4}$ &   $1.8167\times 10^{-2}$ &   $6.6520\times 10^{-1}$ &   $1.6910\times 10^{-2}$ \\
   
    & $0.038$ & $0.43$ & $2.1314\times 10^{-5}$ & $2.8925\times 10^{-4}$ &  $-2.4858\times 10^{-4}$ &   $1.9228\times 10^{-2}$ &   $6.3417\times 10^{-1}$ &   $2.0564\times 10^{-1}$ \\ 
    
    & $0.040$ & $0.45$ & $1.5181 \times 10^{-5}$ &  $2.0542\times 10^{-4}$ &   $1.5184 \times 10^{-5}$ &   $2.2968 \times 10^{-2}$ &   $0.6259 \times 10^{-1}$ &   $2.2834 \times 10^{-1}$
    \\ 
    \hline
	\end{tabular}}
	\tablefoot{Polynomial coefficients of the fitting functions Eqs.~(\ref{eq:IGIMF_Halpha_SFR_fit}), (\ref{eq:IGIMF_Kennicutt_SFR_fit}), and (\ref{eq:IGIMF_Kennicutt_SFR_fit_2}) to the $\psi$--$L_{\mathrm{H\alpha}}$ and $\Phi-L_{\mathrm{H\alpha}}$, $\psi_{\mathrm{K98}}-\psi_{\mathrm{IGIMF}}$ relations, respectively, for different metallicities. The fitting relations of $\psi$--$L_{\mathrm{H\alpha}}$ and $\Phi-L_{\mathrm{H\alpha}}$ relations are defined for the H$\alpha$ luminosity range of $-6 < \log_{10}\big(L_{\mathrm{H\alpha}}/(10^{41}\,\mathrm{erg\,s^{-1}})\big) < 4$ and $\psi_{\mathrm{K98}}-\psi_{\mathrm{IGIMF}}$ relations for SFRs derived from the Kennicutt law in the range of $-6 < \log_{10}\big(\psi_{\mathrm{K98}}/(\mathrm{M_{\odot}\,yr^{-1}})\big) < 4$ (Section~\ref{subsec: SFR-Halpha relation}). They cannot be applied beyond these boundaries.}
  \label{tab:fitting_parameters}
\end{table*}

\end{appendix}
    
\end{document}